\documentstyle[aaspp4]{article}
\newcommand{\kms}{km s$^{-1}$}
\newcommand{\zabs}{$z_{\rm abs}$}
\newcommand{\zem}{$z_{\rm em}$}
\newcommand{\lya}{Ly$\alpha$\ }
\newcommand{\lyb}{Ly$\beta$\ }

\slugcomment{To appear in {\it The Astrophysical Journal}, Dec. 20, 2001}
\lefthead{TRIPP et al.}
\righthead{IONIZATION OF INTERGALACTIC \ion{O}{6}}
\begin{document}

\title{The Ionization and Metallicity of the Intervening 
\ion{O}{6} Absorber at z = 0.1212 in the Spectrum of 
H1821+643\altaffilmark{1}}

\author{Todd M. Tripp,\altaffilmark{2} Mark L. 
Giroux,\altaffilmark{3,4} John T. Stocke,\altaffilmark{3} 
Jason Tumlinson,\altaffilmark{3} and William R. 
Oegerle\altaffilmark{5,6}}

\altaffiltext{1}{ Based on observations with the NASA/ESA 
{\it Hubble Space Telescope}, obtained at the Space 
Telescope Science Institute, which is operated by the 
Association of Universities for Research in Astronomy, 
Inc., under NASA contract NAS 5-26555.}

\altaffiltext{2}{Princeton University Observatory, 
Peyton Hall, Princeton, NJ 08544, 
Electronic mail: tripp@astro.princeton.edu}

\altaffiltext{3}{Center for Astrophysics and Space 
Astronomy and Department of Astrophysical and Planetary 
Sciences, University of Colorado, Boulder, CO 80309.}

\altaffiltext{4}{Current address: Department of Physics and 
Astronomy, East Tennessee State University, Johnson City, 
TN 37614}

\altaffiltext{5}{Department of Physics and Astronomy, Johns 
Hopkins University, Baltimore, MD 21218}

\altaffiltext{6}{NASA Goddard Space Flight Center, 
Greenbelt, MD 20771}

\begin{abstract}
We use high-resolution UV spectra of the radio-quiet QSO 
H1821+643 (\zem\ = 0.297), obtained with the Space 
Telescope Imaging Spectrograph (STIS) and the {\it Far 
Ultraviolet Spectroscopic Explorer (FUSE)}, to study the 
ionization and metallicity of an intervening \ion{O}{6} 
absorption line system at \zabs\ = 0.1212. This absorber 
has the following notable properties: (1) Several galaxies 
are close to the sight line at the absorber redshift, 
including an actively star-forming galaxy at a projected 
distance of 144 $h_{75}^{-1}$ kpc. (2) There is a complex 
cluster of \ion{H}{1} \lya absorption lines near the 
\ion{O}{6} redshift, including at least five components 
spread over a velocity range of $\sim$700 \kms . (3) The 
strongest \lya line in the cluster appears to be composed 
of a mildly saturated component with a typical $b-$value 
blended with a remarkably broad component with $b \approx$ 
85 \kms . (4) The \ion{O}{6} absorption is not aligned with 
the strongest (saturated) \ion{H}{1} absorption, but 
instead is 
well-aligned with the very broad component. (5) The only 
detected species (at the $4\sigma$ level) are \ion{O}{6} 
and \ion{H}{1} despite coverage of strong transitions of 
abundant elements (e.g., \ion{C}{2}, \ion{C}{3}, and 
\ion{C}{4}). Based on these constraints, we find that the 
absorption line properties can be produced in collisionally 
ionized gas  with $10^{5.3} \leq T \leq 10^{5.6}$ K and $-
1.8 \leq$ [O/H] $\leq -0.6$. However, we find that 
photoionization is also viable if the pathlength $l$ 
through the absorbing gas is long enough; simple 
photoionization models require 85 $\leq l \leq$ 1900 kpc 
and $-1.1 \leq$ [O/H] $\leq -0.3$. We briefly discuss how 
observations of X-ray absorption lines due to \ion{O}{7} 
and \ion{O}{8} could be used, in principle, to break the 
ionization mechanism degeneracy, and we conclude with some 
comments regarding the nature of \ion{O}{6} absorbers. 
\end{abstract}

\keywords{ intergalactic medium --- quasars: absorption lines --- 
quasars: individual (H1821+643)}

\section{Introduction}

Any successful cosmological model must account for the 
quantity, distribution, and physical state of the baryons 
in the Universe. While this is likely a more tractable 
challenge than understanding the more exotic non-baryonic 
dark matter and dark energy (i.e., the cosmological 
constant), there is a ``missing baryon problem'' which has 
been known for some time. In brief, inventories of the 
well-observed baryons in the nearby Universe (including 
stars, cool neutral gas, and X-ray emitting galaxy cluster 
gas) find total baryon densities, in units of the closure 
density, of $\Omega _{b} = \rho/\rho _{c} \approx$ 
0.006 (assuming $H_{0}$ = 75$h_{75}$ km s$^{-1}$ Mpc$^{-
1}$, see, e.g., Persic \& Salucci 1992; Fukugita et al. 
1998). This is far less than the expected value based on 
Big Bang nucleosynthesis and deuterium measurements (Burles 
\& Tytler 1998): $\Omega _{b} h_{100}^{2}$ = 
0.019$\pm$0.001 ( = 0.034$\pm$0.002 for $h_{100}$ = 0.75). 
Ironically, the baryons have been more readily accounted 
for in the distant Universe than in the low$-z$ Universe: 
at high redshifts, the ``\lya forest'' -- photoionized, 
diffuse intergalactic gas at $T \sim 10^{4}$ K -- appears 
to contain the vast majority of the baryons predicted by 
the deuterium observations (Rauch et al. 1997; Weinberg et 
al. 1997). At low redshifts, the cool \lya forest is less 
dominant but still makes a substantial contribution to the 
baryon inventory. However, even including recent estimates 
of the baryonic content of the low$-$z Ly$\alpha$ forest 
(Shull, Stocke, \& Penton 1996; Penton, Shull, \& Stocke 
2000), the census of baryons at the present epoch falls 
well short of the expected $\Omega _{b}$. Consequently, 
cosmologists are confronted with an important question: 
where are the missing baryons at the present epoch?

A promising possible answer to this question has been 
provided by cosmological simulations of the growth of large 
scale structure. According to hydrodynamic simulations of 
structure formation, at the present epoch 30$-$50\% of the 
baryons (by mass) are located in low-density, intergalactic 
gas which has been shock-heated to $10^{5} - 10^{7}$ K (Cen 
\& Ostriker 1999a; Dav\'{e} et al. 1999, 2001). Following 
Cen \& Ostriker, we refer to this $10^{5} - 10^{7}$ K gas 
as the warm/hot intergalactic medium to distinguish it from 
the hotter gas observed in rich galaxy clusters. Such gas 
is challenging to observe. Its X-ray emission is extremely 
difficult to detect, especially at lower energies where 
confusion due to foreground emission and absorption is 
substantial. Current X-ray observations do not exclude this 
hot IGM baryon reservoir (Kuntz, Snowden, \& Mushotzky 
2001; Phillips, Ostriker, \& Cen 2001), and there are some 
indications of large-scale filaments of diffuse gas 
delineated by X-ray emission (e.g., Scharf et al. 2000; 
Rines et al. 2001). However, Voit, Evrard, \& Bryan (2001) 
and Croft et al. (2001) have demonstrated that it is 
difficult to distinguish between X-ray emission from 
diffuse intergalactic gas outside of virialized structures 
(as predicted by the cosmological simulations) and X-ray 
emission from virialized groups and clusters. The gas may 
be a significant baryon repository in either case, but the 
distinction has important physical implications, e.g., 
regarding the relative importance of gravitational vs. 
nongravitational heating processes.

QSO absorption lines provide an alternative means to test 
the prediction that a significant quantity of baryons are 
in the warm/hot IGM. Absorption lines such as the 
\ion{O}{6} $\lambda \lambda$1031.9,1037.6 and \ion{Ne}{8} 
$\lambda \lambda$ 770.4,780.3 doublets provide sensitive 
probes of $10^{5} - 10^{6}$ K gas in collisional ionization 
equilibrium (Verner, Tytler, \& Barthel 1994). Searches for 
the \ion{O}{6} absorption lines at \zabs $\lesssim$ 0.3 
with the Space Telescope Imaging Spectrograph (STIS) on 
board the {\it Hubble Space Telescope (HST)} have revealed 
that the number of intervening \ion{O}{6} absorbers per 
unit redshift ($dN/dz$) is remarkably high (Tripp, Savage, 
\& Jenkins 2000; Tripp \& Savage 2000). 
Furthermore, with reasonable assumptions about the 
metallicity and ionization of the gas, these studies have 
shown that \ion{O}{6} absorption systems are probably an 
important baryon reservoir.

Since the \ion{O}{6} ion fraction peaks at $T \sim 3 \times 
10^{5}$ K in collisional ionization equilibrium, these 
\ion{O}{6} absorption studies provide tantalizing evidence 
that the warm/hot IGM is an important baryon repository at 
$z \sim 0$. However, there are several issues which must be 
addressed in order to build a compelling case. First, the 
number of \ion{O}{6} systems (and sight lines) in the 
published STIS observations is rather small, and 
consequently quantities such as $dN/dz$ and the 
cosmological mass density have large uncertainties. More 
observations are needed, and new programs to build the 
sample are underway. Second, the \ion{O}{6} lines could 
arise in gas which is not collisionally ionized, but rather 
is photoionized by the UV background from QSOs and active 
galaxies. Or, the gas could be collisionally ionized but 
out of equilibrium, e.g., because it is able to cool more 
rapidly than it can recombine. Photo- or non-equilibrium 
ionization is favored in some of the \ion{O}{6} systems 
because they have narrow associated \ion{H}{1} lines (e.g., 
Tripp \& Savage 2000; Savage et al. 2001) which would not 
arise in equilibrium collisional ionization. Alternatively, 
these could be multiphase absorbers. Of course, in some 
cases photoionization and collisional ionization could be 
comparably important. If the gas is photoionized, then 
there are concerns about double-counting in the baryon 
inventory (see below). Third, the metallicity of the 
absorbers is poorly constrained. If these \ion{O}{6} 
systems originate in pockets of high metallicity gas, then 
there is less hydrogen associated with the absorbers and 
their baryonic content is lower [$\Omega_{b}$(\ion{O}{6}) 
$\propto$ (O/H)$^{-1}$]. Finally, the environment in which 
these absorbers are located must be scrutinized. Are these 
absorbers found in unvirialized galaxy filaments or even 
galaxy voids, or do they arise in virialized structures? 
Are these systems due to gas which is gravitationally 
shock-heated when gas accretes onto 
large-scale structures, or could they be heated by 
nongravitational processes such as supernova-driven winds?

In this paper we are primarily interested in the ionization 
and metallicity of the gas. Specifically, we use 
high-resolution UV spectra obtained with STIS and the {\it 
Far Ultraviolet Spectroscopic Explorer (FUSE)} to study the 
ionization and nature of a particular absorption system at 
\zabs\ = 0.1212 in the spectrum of the radio-quiet QSO 
H1821+643 (\zem\ = 0.297). While some \ion{O}{6} absorbers 
are apparently ``intrinsic'' (i.e., close to the QSO 
itself, see Hamann \& Ferland 1999), the H1821+643 absorber 
at \zabs\ = 0.1212 is certainly an intervening system: it 
is highly displaced from the QSO redshift ($\Delta v 
\approx$ 43,400 \kms ) and is located at a redshift where 
there are several galaxies close to the line of sight, 
including a luminous galaxy at a projected distance of 144 
$h_{75}^{-1}$ kpc which shows [\ion{O}{2}], [\ion{O}{3}], 
and H$\beta$ emission lines indicative of active star 
formation (Tripp, Lu, \& Savage 1998; Bowen, Pettini, \& 
Boyle 1998). The ionization mechanism is a crucial issue 
because if the \ion{O}{6} absorbers are photoionized and 
cool rather than collisionally ionized and hot, then they 
still represent a substantial baryon reservoir (see Tripp 
\& Savage 2000), but in this case they may reveal the same 
baryons counted in \lya studies such as Penton et al. 
(2000) [note that all intervening \ion{O}{6} systems that 
we have detected so far are also detected in the \ion{H}{1} 
\lya transition]. Apart from the baryon census, the 
metallicity and physical conditions of the gas are 
interesting in their own right.

The paper is organized as follows. The STIS and {\it FUSE} 
observations and data reduction are described in \S 2. 
Section 3 summarizes the absorption line measurement 
techniques. Constraints are placed on the physical 
conditions and metallicity of the gas in \S 4 including a 
comparison of the absorber properties required in 
collisionally ionized and photoionized scenarios and some 
comments on X-ray absorption lines of \ion{O}{7} and 
\ion{O}{8}. We discuss the results and summarize our 
conclusions in \S 5. Throughout this paper we assume 
$H_{0}$ = 75 \kms\ Mpc$^{-1}$ and $q_{0}$ = 0.0 unless 
otherwise indicated. We also report heliocentric 
wavelengths and redshifts; in this direction, $v_{\rm LSR} 
= v_{\rm helio} + 16$ \kms\ assuming the standard 
definition of the local standard of rest (Kerr \& 
Lynden-Bell 1986).

\section{Observations and Data Reduction}

This paper makes use of observations of H1821+643 taken 
with STIS in both the echelle and first-order grating 
modes, as well as shorter-wavelength observations made with 
the {\it FUSE} satellite. In this section we briefly 
describe the observations and data reduction; see Woodgate 
et al. (1998) and Kimble et al. (1998) for information on 
the design and performance of STIS, and Moos et al. (2000) 
and Sahnow et al. (2000) regarding the {\it FUSE} design 
and performance.

\subsection{STIS Spectroscopy}

H1821+643 was observed with the medium resolution FUV 
echelle mode (E140M) of STIS on 1999 June 25 and 2000 March 
31 for a total integration time of 50.93 ksec. This STIS 
mode uses the FUV MAMA detector, which generally has very 
low dark background counts. While this detector also has an 
amoebic region that shows elevated dark counts when the 
detector temperature increases (see Figure 1 in Brown et 
al. 2000), the dark count rate was small compared to the 
source count rate for all of the observations employed 
here. The observations made use of the $0\farcs 2 \times 
0\farcs 06$ slit resulting in a line spread function with 
minimal wings (see Figure 13.87 in the STIS Instrument 
Handbook, v4.1). This STIS echelle mode/slit combination 
provides a resolution of $R \ = \ \lambda /\Delta \lambda 
\approx$ 46,000 (FWHM $\approx$ 7 \kms , see Kimble et al. 
1998) and wavelength coverage from $\sim$1150 to 1710 \AA\ 
with four small gaps between orders at $\lambda >$ 1630 \AA 
. Several STIS first-order grating observations of 
H1821+643 were also obtained on 1999 June 24. Most relevant 
to this paper is the STIS G230M observation extending from 
1724 to 1814 \AA , which covers the \ion{C}{4} doublet at 
\zabs\ = 0.1212. This first-order grating spectrum has a 
resolution of $\sim$30 \kms\ (FWHM).

The data were reduced with the STIS Team version of CALSTIS 
at the Goddard Space Flight Center. The individual echelle 
spectra were flatfielded, extracted, and wavelength and 
flux calibrated with the standard techniques. Then a 
correction for scattered light\footnote{Inspection of the 
cores of strongly saturated lines indicates that while the 
scattered light correction is usually quite effective, 
occasionally errors in the flux zero point at the level of 
a few percent of the continuum are evident. Consequently, 
we have included a term due to the flux zero point 
uncertainty in the overall uncertainties in the STIS 
measurements (see \S 3), and we have allowed the 
profile-fitting code to adjust the zero point by a few 
percent (for the STIS data only) as a free parameter.} was 
applied using the method developed by Bowers et al. (2001), 
and the individual spectra were combined weighted by their 
inverse variances averaged over a large, high S/N region. 
Finally, overlapping regions of adjacent orders were also 
coadded weighted inversely by their variances.\footnote{For 
the overlapping regions of adjacent orders, the weighting 
was determined on a pixel-by-pixel basis. However, a 
five-pixel boxcar smoothing was applied to the error 
vectors for the determination of the weights so that pixels 
with larger noise fluctuations are not inappropriately 
over- or under-weighted.} The first and last ten pixels in 
each order were not coadded to avoid spurious pixel values 
often seen near order edges. 

The STIS G230M first-order grating observation covering the 
\ion{C}{4} $\lambda \lambda$1548.2,1550.8 doublet at \zabs\ 
= 0.1212 was flatfielded and wavelength calibrated in the 
normal way, but the spectrum was extracted with the optimal 
method of Robertson (1986). Two 50-pixel wide regions 
centered 52 pixels away from the spectrum (one on each 
side) were used to determine the background. The two G230M 
spectra were also combined with weighting based on signal-
to-noise. To ensure that the wavelength scale of the G230M 
spectrum is aligned with that of the E140M echelle 
spectrum, the Milky Way \ion{Ni}{2} $\lambda$1741.5 and 
$\lambda$1751.9 lines covered in the G230M spectrum were 
compared to the Milky Way \ion{Ni}{2} $\lambda$1317.2 and 
$\lambda$1370.1 lines recorded in the E140M spectrum. 

\subsection{FUSE Spectroscopy}

H1821+643 was initially observed with {\it FUSE} on 1999 
October 10 and 13; the total integration time was 48.8 
ksec. Results from this observation have been presented by 
Oegerle et al. (2000). Subsequently, the QSO was 
re-observed with {\it FUSE} on 2000 July 24 with a total 
integration time of 62.7 ksec. Both observations used the 
large ($30'' \times 30''$) LWRS aperture. {\it FUSE} has 
four co-aligned telescopes and Rowland spectrographs which 
record spectra on two microchannel plate detectors. Two of 
the optical channels have LiF coatings to cover the 
1000$-$1187 \AA\ range, and the other two channels have SiC 
coatings to cover 905$-$1105 \AA . During the 1999 October 
observation, only the two LiF channels were successfully 
aligned, but high-quality spectra were obtained with all 
four channels in 2000 July. By design, most wavelengths in 
the {\it FUSE} bandpass are recorded by at least two 
channels (see Figure 3 in Sahnow et al. 2000).

The initial (1999) observations were reduced as described 
in Oegerle et al. (2000). The later observations were 
reduced in a similar fashion with CALFUSE version 1.7.6. 
The data were recorded in time-tag mode and were processed 
to correct for Doppler shifts due to orbital motion, 
subtract backgrounds, and wavelength and flux calibrate the 
spectra. Flatfield and astigmatism corrections were not 
sufficiently tested at the time of processing and were not 
applied. CALFUSE assumes that the background is uniform 
across a given detector, which is an adequate approximation 
for an object as bright as H1821+643. Oegerle et al. (2000) 
report that the resolution of the 1999 October spectrum is 
20$-$25 \kms . The resolution of later observations are 
somewhat better; the FUSE PI Team finds that the current 
resolution ranges from $\sim$17 to 25 \kms\ (e.g., Savage 
et al. 2001). In both cases the resolution depends on the 
wavelength and channel used to record the data. Because of 
the varying resolution of data from different channels and 
observation dates, there are differing views on how to best 
make use of multiple detections of a particular line. We 
discuss our strategy regarding this issue in \S 3.

While the wavelength scale of the 1999 October data can 
show large errors (10-30 \kms ) over small intervals, this 
problem was subsequently corrected and is not evident in 
the 2000 July data. The relative dispersion solution of the 
2000 July data is accurate to $\sim$6 \kms . However, there 
can still be an error in the zero point of the wavelength 
scale, e. g., due to imperfect centering of the target in 
the aperture. To set the zero point of the {\it FUSE} data, 
the multicomponent Milky Way \ion{Fe}{2} $\lambda$1144.9 
line was compared to the Galactic \ion{Fe}{2} 1608.5 line 
recorded in the STIS E140M spectrum. Similarly, Lyman 
series lines in the {\it FUSE} spectrum at \zabs\ = 0.225, 
which show distinctive component structure, were compared 
to the Ly$\gamma$ line at this redshift in the STIS 
spectrum. As in Oegerle et al. (2000), the wavelength scale 
of the 1999 October data was corrected on a line-by-line 
basis by comparison to appropriate analogous lines in the 
STIS E140M observation.

\section{Absorption Line Measurements}

\begin{figure}
\plotone{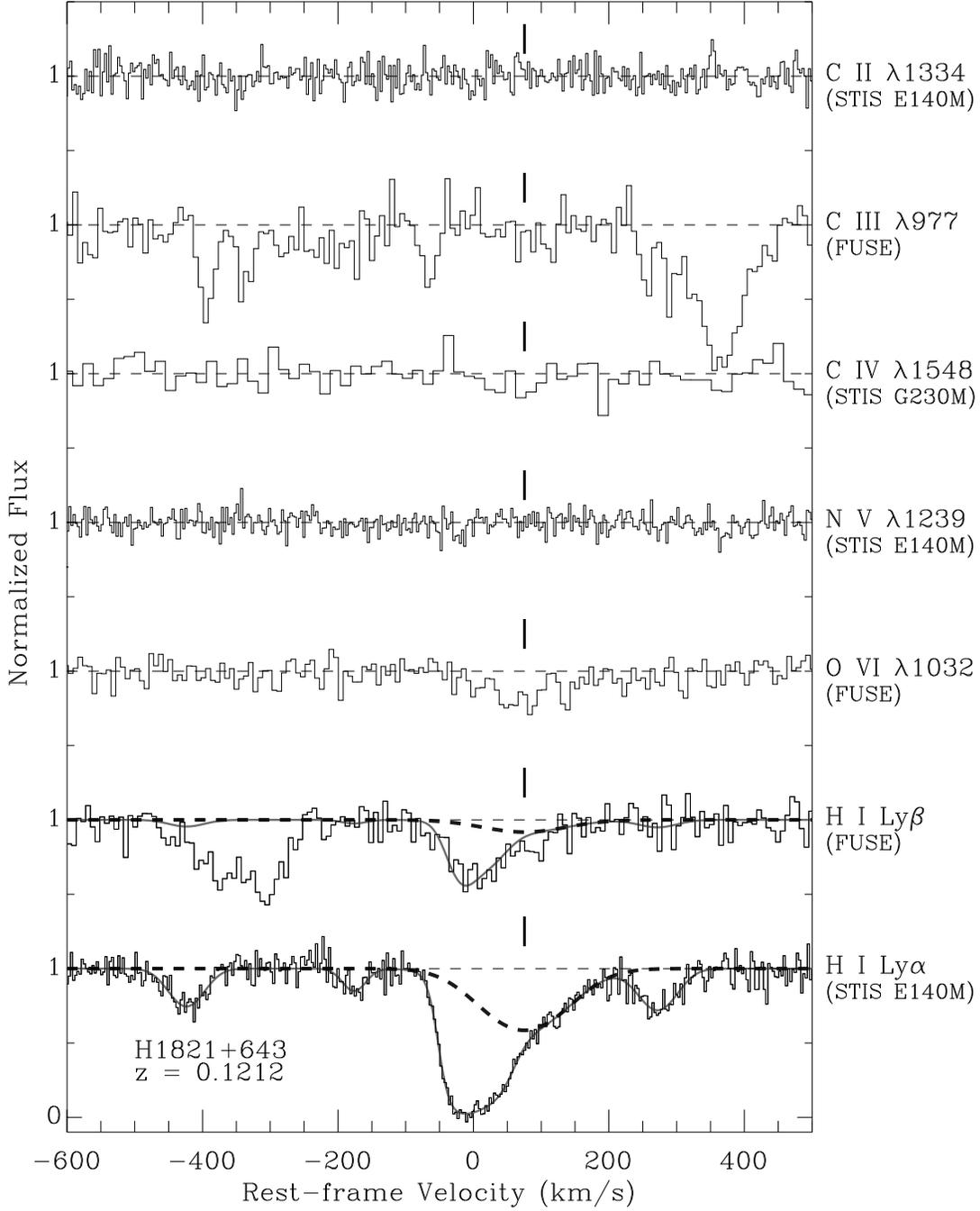}
\caption[]{\footnotesize Continuum-normalized absorption profiles of the 
\ion{H}{1} Ly$\alpha$, Ly$\beta$, and \ion{O}{6} $\lambda$ 
1032 lines detected at \zabs\ = 0.1212 in the spectrum of 
H1821+643 (lower three profiles); the profiles are plotted 
vs. rest-frame velocity where $v =$ 0 at \zabs\ = 0.1212. 
The \ion{C}{2} $\lambda 1334.5$, \ion{C}{3} $\lambda 
977.0$, \ion{C}{4} $\lambda 1548.2$, and \ion{N}{5} 
$\lambda 1238.8$ lines are not significantly detected at 
this redshift, and the spectral regions of these undetected 
lines are shown in the upper portion of the figure. The 
profile fit which yields the component parameters in 
Table~\ref{compprop} is overplotted on the Ly$\alpha$ and 
Ly$\beta$ profiles with a solid line, and the broad 
\ion{H}{1} component discussed in \S 4 is shown with a 
dashed line. For reference, the velocity centroid of this 
broad component is indicated with a tick mark above each 
profile. Note that there are several unrelated lines near 
the \ion{C}{3} and Ly$\beta$ profiles.\label{stack}}
\end{figure}

The absorption profiles of detected lines at \zabs\ = 
0.1212 as well as regions of undetected species of interest 
are shown in Figure~\ref{stack}. Standard processing of 
{\it FUSE} data provides highly oversampled spectra, so we 
have rebinned the {\it FUSE} data to $\approx$ 7 \kms\ 
pixels. The STIS data are optimally sampled, so no 
rebinning has been applied. Remarkably, despite the close 
proximity of several galaxies (see \S 1) and a relatively 
strong Ly$\alpha$ line, only Ly$\alpha$, Ly$\beta$, and the 
\ion{O}{6} $\lambda$1032 transitions are clearly detected 
at the 4$\sigma$ level or better at this redshift (we argue 
below that there is evidence of \ion{O}{6} $\lambda$1038 
absorption as well, although the transition is blended with 
an unrelated line). Table~\ref{lineprop} lists rest-frame 
equivalent widths ($W_{r}$) of the detected absorption 
lines at \zabs\ = 0.1212, measured using the methods of 
Sembach \& Savage (1992), as well as $4\sigma$ upper limits 
on undetected lines. Note that contributions due to the 
uncertainty in the curvature and height of the continuum as 
well as a 2\% uncertainty in the flux zero point are 
included with the statistical uncertainties in the overall 
errors in Table~\ref{lineprop}. Integrated apparent column 
densities (Savage \& Sembach 1991), which similarly include 
continuum placement and zero point uncertainties, are also 
provided in Table~\ref{lineprop}.

The quantities in Table~\ref{lineprop} are integrated over 
the full velocity range of the main \lya feature but do not 
include the three high-velocity weak \lya lines evident in 
Figure~\ref{stack} at $v = -425, -179,$ and +270 \kms . 
However, it is obvious from Figure~\ref{stack} that even 
the main \lya feature is composed of at least two 
substantially blended components. To deblend these 
components and measure line widths, we have used the Voigt 
profile-fitting software of Fitzpatrick \& Spitzer (1997) 
to simultaneously fit the \lya and Ly$\beta$ profiles. For 
the STIS echelle data (Ly$\alpha$), we have accounted for 
instrumental broadening using the line spread functions 
from the STIS handbook; for FUSE data (Ly$\beta$), we have 
assumed that the LSF is adequately described by a Gaussian 
with FWHM $\approx$ 20 \kms . The profile-fitting results 
are summarized in Table~\ref{compprop}. The final model 
profile is overplotted on the Ly$\alpha$ and Ly$\beta$ 
profiles in Figure~\ref{stack} with a solid line.

\begin{deluxetable}{lccccc}
\tablewidth{0pc}
\tablecaption{Equivalent Widths and Integrated Column Densities
 of the \ion{O}{6} Absorber at z = 0.1212\label{lineprop}}
\tablehead{Species & $\lambda _{0}$\tablenotemark{a} & 
$f$\tablenotemark{a} & $W_{\rm r}$\tablenotemark{b} & log $N_{\rm 
a}$\tablenotemark{c} & Spectrograph \\
 \ & (\AA ) & \ & (m\AA ) & \ & \ }
\startdata
\ion{H}{1}....... & 1215.67 & 0.416 & 595$\pm$12 & 
$>$14.38\tablenotemark{d} & STIS E140M \\
\ion{H}{1}....... & 1025.72 & 0.0791 & 158$\pm$14 & 
14.43$\pm$0.04 & FUSE\tablenotemark{e} \\
\ion{O}{6}......  & 1031.93 & 0.133 & 90$\pm$17 & 
14.02$\pm$0.07 & FUSE\tablenotemark{f} \\
\ion{N}{5}....... & 1238.82 & 0.156 & 
$<$56\tablenotemark{g} & $<$13.42\tablenotemark{g} & STIS 
E140M \\
\ion{C}{4}......  & 1548.20 & 0.191 & 
$<$144\tablenotemark{g} & $<$13.55\tablenotemark{g} & STIS 
G230M \\
\ion{C}{3}......  & 977.02 & 0.759 & 
$<$104\tablenotemark{g} & $<$13.21\tablenotemark{g} & 
FUSE\tablenotemark{h} \\
\ion{C}{2}....... & 1334.53 & 0.127 & 
$<$62\tablenotemark{g} & $<$13.49\tablenotemark{g} & STIS 
E140M \\
\ion{Si}{4}...... & 1393.76 & 0.514 & 
$<$55\tablenotemark{g} & $<$12.79\tablenotemark{g} & STIS 
E140M \\
\ion{Si}{3}...... & 1206.50 & 1.669 & 
$<$70\tablenotemark{g} & $<$12.51\tablenotemark{g} & STIS 
E140M
\enddata
\tablenotetext{a}{Rest frame vacuum wavelength and 
oscillator strength from Morton (2001) or Morton (1991).}
\tablenotetext{b}{Rest frame equivalent width integrated 
from $-$110 to 210 \kms\ where $v =$ 0 \kms\ at \zabs\ = 
0.1212.}
\tablenotetext{c}{Apparent column density integrated from 
$-$110 to 210 \kms . For undetected lines, 4$\sigma$ upper 
limits are derived from the upper limits on $W_{\rm r}$ 
using the linear portion of the curve of growth.}
\tablenotetext{d}{Saturated absorption line.}
\tablenotetext{e}{Weighted mean of individual measurements 
from the observations with the LiF1b and LiF2a detectors on 
1999 October 17 and 2000 July 24.}
\tablenotetext{f}{ Weighted mean of individual measurements 
from the observations with the LiF1b and LiF2a detectors on 
1999 October 17 and the LiF2a detector on 2000 July 24. The 
LiF1b spectrum from 2000 July 24 was adversely affected by 
the vertical stray light ``stripe'' (Sahnow et al. 2000) 
and is too noisy to provide a useful measurement.}
\tablenotetext{g}{4$\sigma$ upper limit.}
\tablenotetext{h}{Based on the LiF2a spectrum from 2000 
July 24 only.}
\end{deluxetable}

\begin{deluxetable}{lccc}
\tablewidth{0pc}
\tablecaption{Component Parameters of the Ly$\alpha$ and 
Ly$\beta$ 
Absorption Lines at $z$ = 0.1212\tablenotemark{a} 
\label{compprop}}
\tablehead{Redshift\tablenotemark{b} & $v$\tablenotemark{c} 
& $b$ (km s$^{-1}$) & log $N$}
\startdata
0.11961 & $-$425 & 35$\pm$9         & 13.14$\pm$0.04 \\
0.12053 & $-$179 & 23$^{+15}_{-9}$  & 12.71$\pm$0.08 \\
0.12112 & $-$21  & 26$^{+13}_{-9}$  & 13.93$\pm$0.37 \\
0.12125 & 13     & 40$^{+44}_{-21}$ & 14.04$\pm$0.36 \\
0.12147 & 72     & 85$^{+37}_{-26}$ & 13.78$\pm$0.17 \\
0.12221 & 270    & 38$\pm$9         & 13.22$\pm$0.04 
\enddata
\tablenotetext{a}{Based on simultaneous profile fitting of 
the Ly$\alpha$ and Ly$\beta$ lines with the software of 
Fitzpatrick \& Spitzer (1997) and the STIS E140M line 
spread functions from the STIS Instrument Handbook. The 
FUSE LSF was taken to be a Gaussian with FWHM $\approx$ 20 
\kms .}
\tablenotetext{b}{Heliocentric redshift; the conversion 
from a heliocentric velocity scale to the Local Standard of 
Rest velocity scale is given by $v_{\rm LSR} = v_{\rm 
helio} + 16$ \kms\ assuming the Sun is moving in the 
direction $l = 56^{\circ}, b = 23^{\circ}$ at 19.5 \kms\ 
(Kerr \& Lynden-Bell 1986).}
\tablenotetext{c}{Component velocity centroid in the rest 
frame of the absorption system where $v$ = 0 \kms\ at 
\zabs\ = 0.1212.}
\end{deluxetable}

Two comments on the profile-fitting results are in order. 
First, the Ly$\beta$ profile shows excess absorption 
compared to the model profile over several pixels at $v 
\approx$ 90 \kms . This is due to noise. Since the \lya $f-
$value is a factor of 5 greater than the \lyb $f-$value, 
and since the equivalent width goes as $f\lambda ^{2}$, 
this ``component'' would be very prominent in the \lya 
profile if it were real. We have attempted to find a set of 
components which produce this extra \lyb absorption without 
violating the constraints set by the \lya profile, and we 
have been unsuccessful. Since {\it FUSE} is known to have 
complex and substantial fixed-pattern noise (Sahnow et al. 
2000), it is not too surprising to occasionally encounter a 
noise feature of this sort. Second, the model profile 
fitted to the \ion{H}{1} lines is not unique. The 
smoothness of the \lya profile at $100 \lesssim v \lesssim 
200$ \kms\ provides some support for fitting this wing with 
a single broad component, but a comparably good fit can be 
obtained with several narrower components spread over this 
velocity range.

In the case of the {\it FUSE} data, we are primarily 
interested in {\it total} equivalent widths and column 
densities integrated across the full velocity range of the 
main absorption feature (i.e., $-110 \leq v \leq +210$ 
\kms\ in Figure~\ref{stack}). Given the breadth of the 
absorption features of interest and the moderate 
differences in resolution of the various data, it is 
reasonable to coadd the various spectra to increase the 
S/N. Nevertheless, to be conservative we have measured 
integrated equivalent widths and apparent column densities 
from the individual spectra separately, then we have 
determined the mean of the individual measurements, each 
weighted inversely by its variance. These weighted means 
are the final integrated quantities reported in 
Table~\ref{lineprop}. For profile fitting, on the other 
hand, we fitted the STIS \lya and {\it FUSE} Ly$\beta$ 
profiles simultaneously. Consequently, to make full use of 
the {\it FUSE} observations, we first coadded Ly$\beta$ 
profiles with the usual weighting based on the inverse 
variances of the individual spectra. We similarly coadded 
the {\it FUSE} \ion{O}{6} data for the purposes of the next 
paragraph. Note that Figure~\ref{stack} shows these coadded 
Ly$\beta$ and \ion{O}{6} profiles. Since the Ly$\beta$ 
lines are detected at high significance in the individual 
spectra, we elected to coadd only the Ly$\beta$ profiles 
from the two LiF channels obtained in 2000 July and thereby 
preserve the resolution as much as possible. However, the 
\ion{O}{6} lines are detected at lower significance in the 
individual channels, and one of the LiF channels from 2000 
July is not useful in the \ion{O}{6} region due to the 
vertical stray light stripe caused by scattered terrestrial 
\lya emission during spacecraft orbital daytime (see \S 2.3 
in Sahnow et al. 2000). Consequently, for the \ion{O}{6} 
lines we have coadded both of the LiF channels from the 
1999 observations with the one useful LiF channel (LiF2) 
from the 2000 observations. We note hat the measurements 
from the individual channels are consistent within their 
1$\sigma$ uncertainties.

\begin{figure}
\plotone{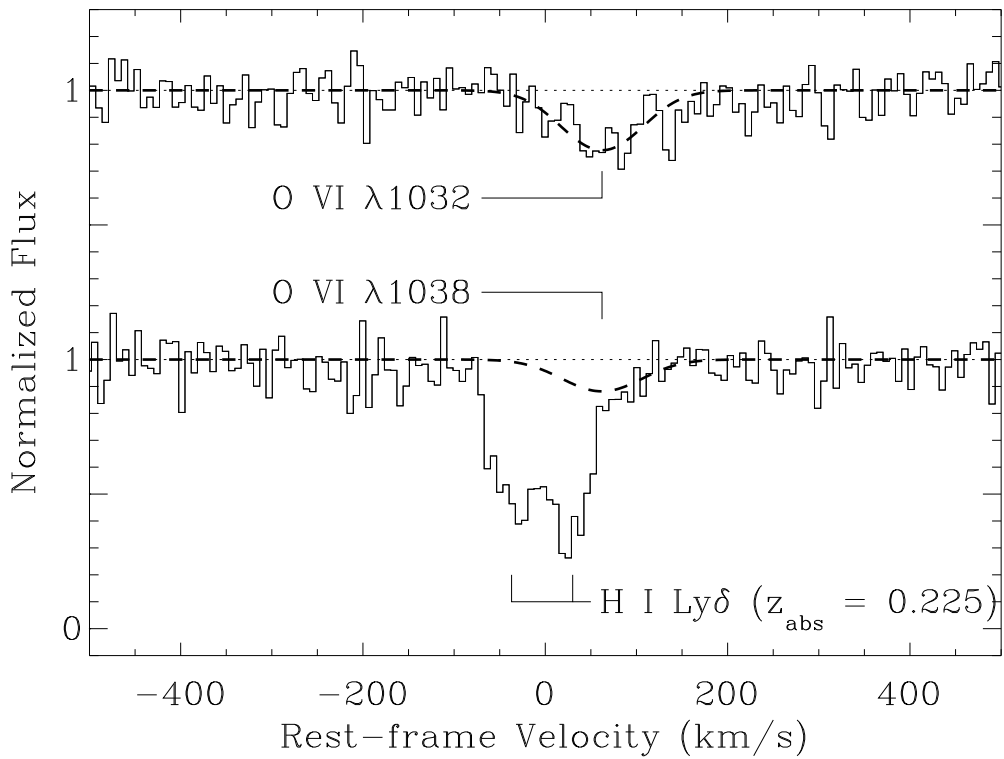}
\caption[]{Continuum-normalized absorption profile of the 
\ion{O}{6} $\lambda$1032 line at \zabs\ = 0.1212 (upper 
histogram), and the spectral region where the corresponding 
\ion{O}{6} $\lambda$1038 line is expected (lower 
histogram), both plotted versus rest-frame velocity where 
$v$ = 0 \kms\ at \zabs\ = 0.1212. Both profiles are derived 
from the coadded data from different channels and 
observation dates (see text). The \ion{O}{6} $\lambda$1038 
line is strongly blended with a Ly$\delta$ line at \zabs\ = 
0.225.  Nevertheless, there is evidence that the \ion{O}{6} 
$\lambda$1038 line is present. The dashed line overplotted 
on the $\lambda$1032 profile shows the result of a 
single-component fit to the 1032 \AA\ line {\it only}. The 
velocity, Doppler parameter, and column density from this 
fit were used to predict the strength of the corresponding 
(blended) $\lambda$1038 transition, and the predicted 1038 
\AA\ line is overplotted with a dashed line on the lower 
profile. An absorption feature with the predicted velocity, 
stength, and width is readily apparent.\label{oviblend}}
\end{figure}

Since \ion{O}{6} has a resonance-line doublet, its 
absorption lines can usually be unambiguously identified. 
Unfortunately, the weaker 1038 \AA\ line of the \ion{O}{6} 
doublet at \zabs\ = 0.1212 in the spectrum of H1821+643 is 
strongly blended with the Ly$\delta$ line from the 
absorption system at \zabs\ = 0.225, which may lead some 
readers to question the reliability of the identification 
of the stronger \ion{O}{6} $\lambda$1032 line shown in 
Figure~\ref{stack}. However, since the Ly$\delta$ line is 
not strongly saturated, we should be able to demonstrate 
that the 1038 \AA\ line is present if the S/N is 
sufficient, and indeed we do find evidence of the 1038 \AA\ 
line at the expected wavelength and strength. 
Figure~\ref{oviblend} shows this evidence. This figure 
shows the continuum-normalized, coadded \ion{O}{6} 
profiles. A single-component fit to the \ion{O}{6} 
$\lambda$1032 line {\it only} is shown with a dashed line 
overplotted on the 1032 \AA\ profile (upper histogram). The 
velocity, column density, and Doppler parameter from this 
fit predict a 1038 \AA\ absorption line shown with a dashed 
line shown overplotted on the lower histogram in 
Figure~\ref{oviblend}. Absorption is clearly present in the 
lower profile at the expected velocity with the expected 
strength.  This gives us confidence that the \ion{O}{6} 
identification is correct.

\section{Physical Conditions and Metallicity}

In this section we explore the implications of the 
absorption-line measurements regarding the ionization and 
metallicity of the gas. We first consider collisional 
ionization, and then we examine photoionization. Before 
assessing the ionization mechanism, we comment on several 
interesting features of this absorber: 

\begin{figure}
\plotone{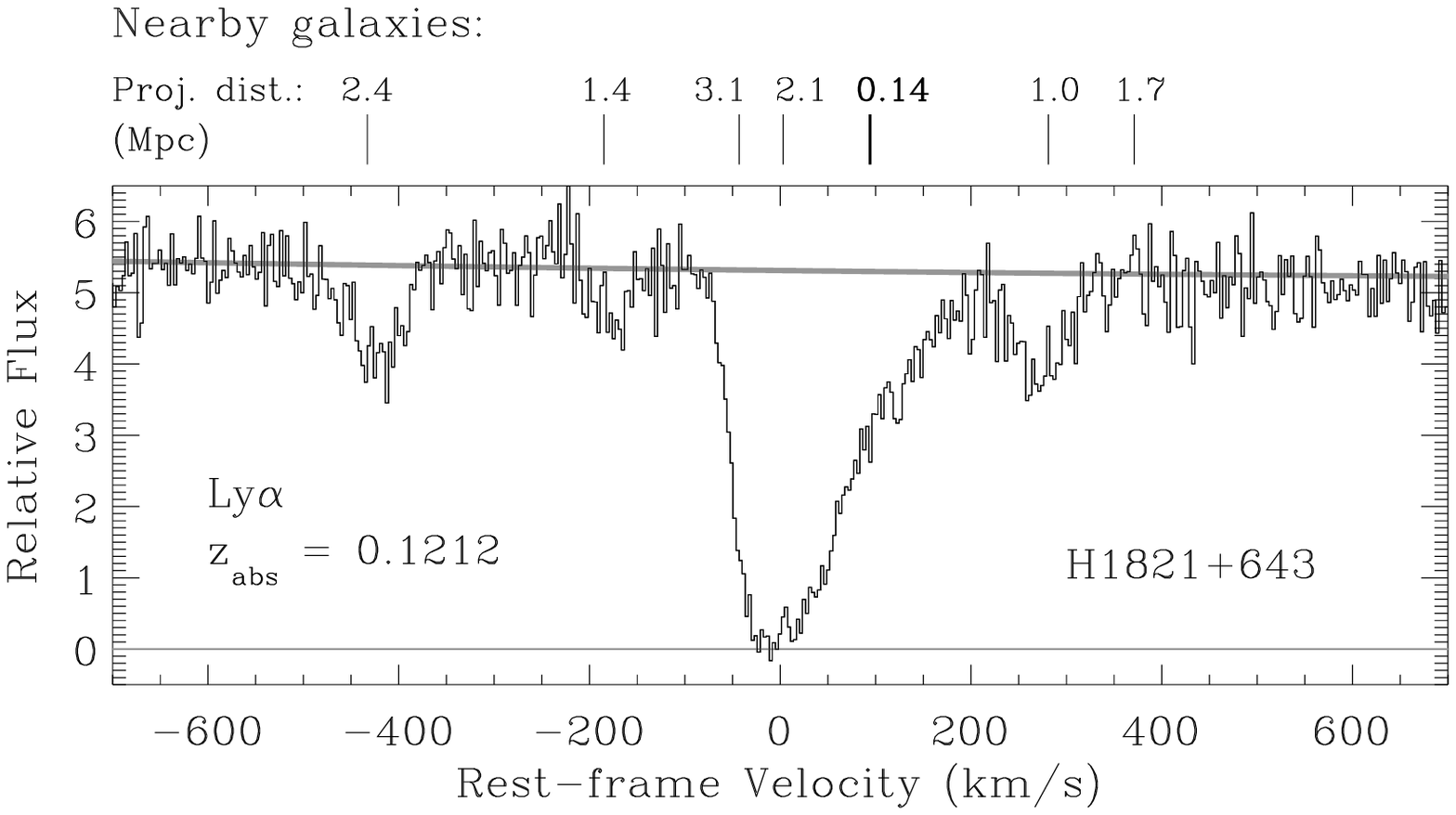}
\caption[]{Cluster of \lya absorption lines near the 
redshift of the \ion{O}{6} absorber at \zabs\ = 0.1212. The 
relative (unnormalized) flux is plotted versus 
rest-frame velocity where $v = 0$ \kms\ at \zabs\ = 0.1212, 
and the continuum placement used for the measurements in 
Tables~\ref{lineprop} and \ref{compprop} is shown with a 
solid gray line. Seven galaxies with measured redshifts are 
close to the line of sight at $z \approx 0.1212$; their 
velocities in this frame [$\Delta v = c\Delta z/(1 + z)$] 
are indicated with tick marks at the top of the panel, and 
the numbers above each tick indicate the projected distance 
of the galaxy from the sight line. The uncertainties in the 
galaxy redshifts are estimated to be 50-100 \kms 
.\label{lyagal}}
\end{figure}

\begin{enumerate}
\item The \ion{O}{6} absorption occurs at a redshift where 
there is a cluster of \lya absorption lines; at least five 
\lya lines are present within a velocity interval of 
$\sim$700 \kms\ (see Figures \ref{stack} and \ref{lyagal}).

\item Furthermore, there are several galaxies within 
$\pm$500 \kms\ of this redshift with projected distances 
from the sight line ranging from 144 $h_{75}^{-1}$ kpc to 
3.1 Mpc (see Tripp et al. 1998). Figure~\ref{lyagal} shows 
the velocities and projected distances of these galaxies 
with respect to the \lya absorption lines [Tripp et al. 
(1998) estimate that the galaxy redshift uncertainties are 
$\sim 50-100$ \kms ]. As noted in \S 1, the closest galaxy 
has an emission line spectrum which suggests that it is 
actively forming stars.

\item In addition to an ordinary, mildly saturated narrow 
line, the main \lya profile appears to contain a relatively 
broad component at $v =$ 72 \kms\ with $b \approx$ 85 \kms\ 
(see Table 2; this broad component is shown with a heavy 
dashed line in Figure~\ref{stack}). This implies an upper 
limit on the temperature of the gas at this $v$: $T \leq 
mb^{2}/2k = 4.3 \times 10^{5}$ K. Therefore this absorption 
could arise in the warm/hot, shock-heated phase predicted 
by cosmological simulations. This is an upper limit because 
the profile could be broadened by factors other than 
thermal motions such as turbulence, multiple blended 
components, or expansion of the Universe. If the broadening 
is predominantly due to cosmological expansion, then the 
implied pathlength $l$ through the absorber is given by 
$\Delta v \approx 15.0$ \kms\ ($l$/200 kpc)$h_{75}$. Taking 
$\Delta v \approx$ FWHM = 142 \kms\ for the broad 
\ion{H}{1} component, we obtain $l \lesssim$ 1.9 $h_{75}^{-
1}$ Mpc in this case.

\item The \ion{O}{6} absorption is not aligned with the 
strongest portion of the \ion{H}{1} profile (i.e., the 
saturated component at $v \approx 0$ \kms ). Rather, the 
\ion{O}{6} is reasonably well-aligned with the broad 
\ion{H}{1} component at $v =$ 72 \kms .\footnote{Oegerle et 
al. (2000) suggested that the H~I Ly$\beta$ and 
O~VI centroids were offset but could not be sure of 
the magnitude of the offset because of problems with the 
wavelength scale in the early {\it FUSE} data. These 
problems have been rectified, and the offset is now 
reliably determined.} The \ion{O}{6} line is also 
apparently broad; the profile fit shown in 
Figure~\ref{oviblend} has $b = 59^{+30}_{-20}$ \kms . This 
would imply a high temperature if entirely due to thermal 
motions and may be evidence of another source of broadening 
(see below). The upper limit on the temperature of the 
\ion{O}{6} gas for $b = 59$ \kms\ is $T \leq 3.3 \times 
10^{6}$ K.

\item As noted above, no metals apart from \ion{O}{6} are 
significantly detected despite coverage of very strong 
resonance transitions of abundant elements (e.g., 
\ion{C}{2} $\lambda$1334, \ion{C}{3} $\lambda$977, and 
\ion{Si}{3} $\lambda$1206). There is a marginal feature 
near the expected wavelength of \ion{C}{4} $\lambda$1548 
(see Figure~\ref{stack}), but its significance is less than 
3$\sigma$ and no feature is evident at the expected 
wavelength of \ion{C}{4} $\lambda$1550. Consequently, we 
use the 4$\sigma$ upper limit on \ion{C}{4} for all 
purposes of this paper, and we obtain 
$N$(\ion{O}{6})/$N$(\ion{C}{4}) $\geq$ 3.0.
\end{enumerate}

\subsection{Collisional Ionization}

Can collisional ionization produce the observed properties 
of this \ion{O}{6} absorber? In collisionally ionized gas 
in equilibrium (we briefly discuss non-equilibrium models 
at the end of \S 4.2), the \ion{O}{6} ionization fraction 
is maximized at $T \approx$ 300,000 K (Sutherland \& Dopita 
1993). The \ion{H}{1} ion fraction is small at such 
temperatures, but nevertheless, by virtue of the great 
abundance of hydrogen, \ion{H}{1} \lya absorption 
associated with hot \ion{O}{6} should be detectable. 
Therefore \ion{H}{1} absorption lines which are rather 
broad due to thermal motions are expected to arise in the 
warm/hot IGM discussed by Cen \& Ostriker (1999a), and the 
broad \ion{H}{1} and \ion{O}{6} lines shown in 
Figure~\ref{stack} seem to provide promising evidence for 
the warm/hot gas. Does this hold up under scrutiny? The 
temperature implied by the apparent breadth of the 
\ion{H}{1} component aligned with the \ion{O}{6} is about 
right for collisional ionization. The lower limit on the 
\ion{O}{6}/\ion{C}{4} ratio provides a lower limit on the 
temperature: assuming the gas is in equilibrium, we find 
from the calculations of Sutherland \& Dopita (1993) that 
$N$(\ion{O}{6})/$N$(\ion{C}{4}) $\geq$ 3.0 requires $T \geq 
10^{5.3}$ K. Combined with the temperature upper limit from 
the broad \ion{H}{1} component, we have $10^{5.3} \leq T 
\leq 10^{5.6}$ K.

As noted above, the single-component \ion{O}{6} fit shown 
in Figure~\ref{oviblend} implies $T \leq 3.3 \times 10^{6}$ 
K. While it is possible to detect trace \ion{O}{6} 
absorption in gas with $T \gtrsim 10^{6}$ K, the 
corresponding \ion{H}{1} absorption would be quite broad 
and inconsistent with the observed \lya profile. It seems 
more likely that the \ion{O}{6} is at least partially 
broadened by non-thermal motions or multiple components. 
This non-thermal broadening would affect associated 
\ion{H}{1} as well but would not preclude substantial 
thermal broadening in the \ion{H}{1} profile. For example, 
if we assume that the broad \ion{H}{1} and \ion{O}{6} 
absorption lines arise in the same gas and that the 
non-thermal motions can be adequately described by a 
Gaussian profile, then we can express the $b-$value as 
$b^{2} = b^{2}_{\rm nt} + 2kT/m$ and solve for $T$ and 
$b_{\rm nt}$, the contribution from non-thermal motions. 
With $b$(\ion{H}{1}) = 85 \kms\ and $b$(\ion{O}{6}) = 59 
\kms , we obtain in this way $T = 2.4 \times 10^{5}$ K and 
$b_{\rm nt}$ = 57 \kms .

The ionization mechanism can also be usefully tested by 
considering the metallicity required by the model $-$ a 
model which requires an excessively high metallicity may be 
unrealistic. We estimate the gas metallicity as follows. 
Using the usual logarithmic notation, [O/H] = log(O/H) $-$ 
log(O/H)$_{\odot}$, the oxygen abundance can be expressed 
as 
\begin{equation}
\left[ \frac{\rm O}{\rm H}\right] = {\rm log}\left( 
\frac{N({\rm O \ VI})}{N({\rm H \ I)}}\right) + {\rm 
log}\left( \frac{f({\rm H \ I})}{f({\rm O \ VI})}\right) - 
{\rm log}\left( \frac{\rm O}{\rm H}\right) _{\odot} 
\label{metlim}
\end{equation}
where $f$ is the ion fraction and (O/H)$_{\odot}$ is the 
solar oxygen abundance.\footnote{Throughout this paper we 
adopt the solar abundances reported by Grevesse \& Anders 
(1989) and Grevesse \& Noels (1993). These reference 
abundances have been widely used in recent literature. 
However, use of the recent revisions of the solar oxygen 
abundance (Holweger 2001; Allende Prieto et al. 2001; see 
also Sofia \& Meyer 2001) increases [O/H] by $\sim 0.1-
0.2$dex.} Since $N$(\ion{O}{6}) and $N$(\ion{H}{1}) are 
measured quantities and (O/H)$_{\odot}$ is fixed, [O/H] can 
be estimated by using an ionization model to estimate 
$f$(\ion{H}{1})/$f$(\ion{O}{6}).

\begin{figure}
\plotone{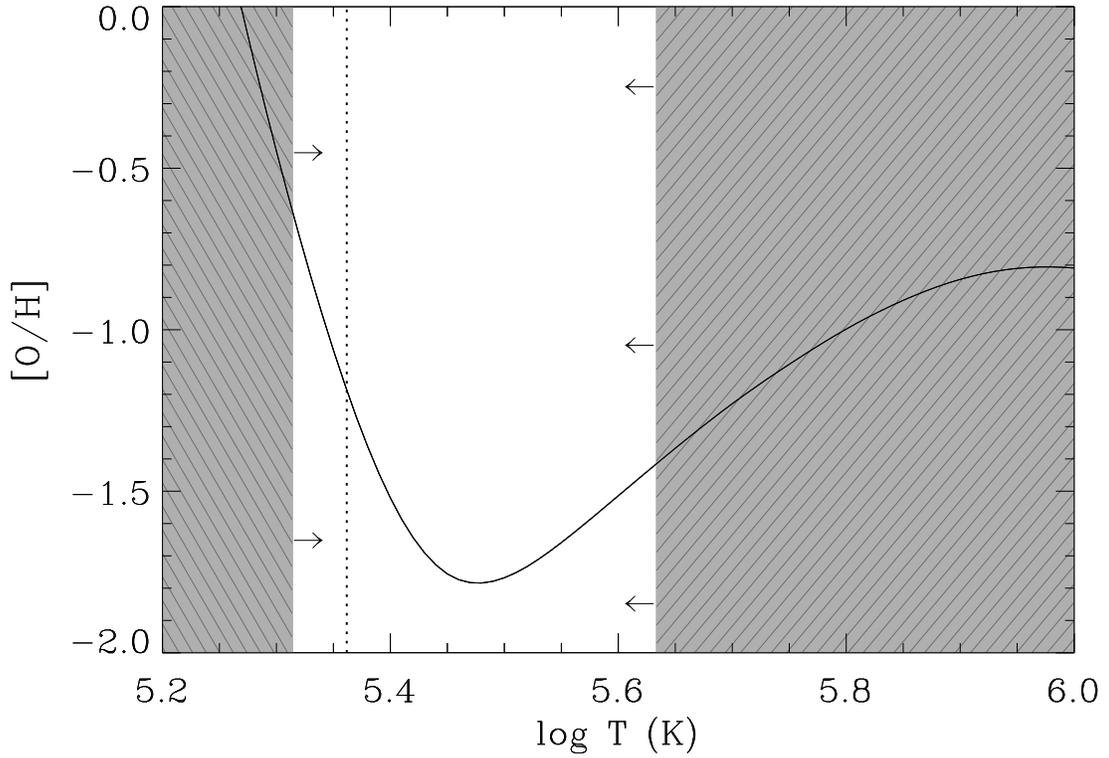}
\caption[]{The logarithmic oxygen abundance [O/H] required 
to produce the observed \ion{O}{6} and \ion{H}{1} column 
densities {\it in the broad component} of the absorber at 
\zabs\ = 0.1212, assuming the gas is collisionally ionized 
and in equilibrium, as a function of temperature (solid 
line). This follows from the \ion{O}{6} and \ion{H}{1} 
ionization fractions vs. $T$ (from Sutherland \& Dopita 
1993) and equation~\ref{metlim}. The gray regions of the 
plot indicate regions which are not allowed by other 
constraints; the lower limit on $T$ is set by the lower 
limit on $N$(\ion{O}{6})/$N$(\ion{C}{4}), and the upper 
limit on $T$ is set by the $b-$value of the broad 
\ion{H}{1} component associated with the \ion{O}{6} 
absorption. The vertical dotted line indicates the mean 
post-shock temperature for a standard shock with $v_{\rm 
s}$ = 130 \kms .\label{colloh}}
\end{figure}

Assuming the broad \ion{H}{1} component shown in 
Figure~\ref{stack} originates in the same gas as the 
\ion{O}{6} absorption, we have log $N$(\ion{H}{1}) = 13.78 
and log $N$(\ion{O}{6}) = 14.02 from Tables~\ref{lineprop} 
and \ref{compprop}. Then, taking 
$f$(\ion{H}{1})/$f$(\ion{O}{6}) vs. $T$ from Sutherland \& 
Dopita (1993), we derive from equation~\ref{metlim} the 
oxygen abundance as a function of $T$ required to match the 
measured $N$(\ion{H}{1}) and $N$(\ion{O}{6}). This is shown 
in Figure~\ref{colloh}. The lower limit on the temperature 
from $N$(\ion{O}{6})/$N$(\ion{C}{4}) and the upper limit on 
$T$ from the \ion{H}{1} line width are also indicated in 
Figure~\ref{colloh}; the hatched gray regions are not 
consistent with these constraints. From this figure, we see 
that within the temperature limits, the observed column 
densities can be reproduced in gas with $-1.8 \lesssim$ 
[O/H] $\lesssim -0.6$, i.e., 0.02 $-$ 0.25 solar 
metallicity.

This allowed range of metallicities is in reasonable 
agreement with theoretical expectations for intergalactic 
gas. For example, the study of IGM metal enrichment by Cen 
\& Ostriker (1999b) indicates that this range of 
metallicity could arise in regions with a wide range of 
overdensity $\delta = \rho /<\rho >$ (see the $z$ = 0 curve 
in their Figure 2), and Z $\sim$ 0.1 Z$_{\odot}$ would not 
be surprising given the \ion{H}{1} column density of the 
absorber studied here. The models of Aguirre et al. (2001), 
which use various semi-analytic prescriptions to explore 
how metals might be distributed within a previously 
computed hydrodynamic simulation, predict somewhat lower 
mean metallicities at $z$ = 0. The mean metallicity is 
somewhat less than 0.02 solar for $\delta <$ 100 in their 
supernova-driven wind models, for example. However, their 
calculation only includes enrichment from galaxies with 
baryonic mass $> 3 \times 10^{10}$ M$_{\odot}$, and 
inclusion of smaller galaxies could increase the 
metallicity by a factor of a few. Furthermore, since this 
is a mean metallicity, there are certainly higher 
metallicity regions in the models, and some of these are 
likely to be consistent with the requirements derived above 
for the \zabs\ = 0.1212 absorber. Of course, it is possible 
that this absorber does not originate in a particularly low 
overdensity region -- the presence of several galaxies 
close to the absorber redshift suggests that this absorber 
may be due to gas within a galaxy group, or at least a 
somewhat overdense filament. However, the metallicity range 
shown in Figure~\ref{colloh} is reasonable in this scenario 
as well: X-ray observations (e.g., Davis, Mulchaey, \& 
Mushotzky 1999; Hwang et al. 1999) of galaxy groups 
indicate that the intragroup medium metallicity is 
consistent with the upper portion of the range derived for 
this absorber.

The \ion{H}{1} and \ion{O}{6} absorption line kinematics 
are intriguing in the context of collisional ionization. 
The juxtaposition of a relatively narrow \ion{H}{1} 
component at $v \approx$ 0 \kms\ with a broad \ion{H}{1} + 
\ion{O}{6} component separated by $\sim$75 \kms\ is 
suggestive of shock-heating in two colliding clouds, with 
the \ion{O}{6} and broad \ion{H}{1} arising in the 
post-shock gas. If we estimate the shock velocity by 
scaling the apparent velocity separation of the components 
by $\sqrt{3}$ (to account for projection effects), we 
estimate that two clouds colliding at a velocity $v_{\rm s} 
= \sqrt{3} \times 75$ \kms\ = 130 \kms\ would be 
shock-heated to $T \approx 2.3 \times 10^{5}$ K, assuming a 
standard shock\footnote{For a standard shock with $\gamma$ 
= 5/3, the mean post-shock temperature is given by $T_{\rm 
s} = 1.38 \times 10^{5} (v_{\rm s}/100$ \kms $)^{2}$ K 
(see, e. g., McKee \& Hollenbach 1980; Draine \& McKee 
1993).} with specific heat ratio $\gamma$ = 5/3. This 
temperature, shown with a vertical dotted line in 
Figure~\ref{colloh}, is consistent with the temperature 
constraints derived above.\footnote{If the scale factor to 
correct for projection effects is less than $\sqrt{3}$, 
then $T_{s}$ could drop below the lower limit on the gas 
temperature set by $N$(O~VI)/$N$(C~IV). 
However, the component of the shock velocity which is 
projected onto the 
line-of-sight could also be larger than 75 \kms , which 
would offset this effect. The point here is that the 
observed absorption lines have approximately the right 
velocities to produce O~VI in a shock.} 

Summarizing this section, we find that the observed 
properties of this absorber can be produced by 
collisionally ionized gas in equilibrium with 5.3 $\leq$ 
log $T \ \leq$ 5.6 and $-1.8 \leq$ [O/H] $\leq -0.6$. These 
properties are consistent with expectations for the IGM as 
well as the intragroup medium in galaxy groups. Equilibrium 
collisional ionization is evidently a viable ionization 
mechanism for this absorber.

\subsection{Photoionization}

Is photoionization also viable for this absorber? To 
address this question we have constructed standard models 
with the photoionization code CLOUDY (Ferland et al. 1998). 
The models adopt the usual assumptions: the absorber has 
constant density and the geometry is plane-parallel, the 
gas is photoionized by the UV background from QSOs and AGNs 
at $z \approx$ 0.12 as calculated by Haardt \& Madau 
(1996),\footnote{We have also computed models using the UV 
background calculated by Shull et al. (1999b), and we 
obtain very similar results.} and the mean intensity at 1 
Rydberg is set to $J_{\nu}$(LL) = $1 \times 10^{-23}$ ergs 
s$^{-1}$ cm$^{-2}$ Hz$^{-1}$ sr$^{-1}$, a value in 
reasonable agreement with observations (e.g., Kulkarni \& 
Fall 1993; Maloney 1993; Vogel et al. 1995; Donahue, 
Aldering, \& Stocke 1995; Tumlinson et al. 1999; Dav\'{e} 
\& Tripp 2001) and theoretical predictions (e.g., Haardt \& 
Madau 1996; Fardal, Giroux, \& Shull 1998; Dav\'{e} et al. 
1999; Shull et al. 1999b). With these assumptions, we 
varied the gas metallicity and ionization parameter $U$ ($= 
n_{\gamma}/n_{\rm H}$ = ionizing photon density/total 
hydrogen number density) to find the parameter ranges 
consistent with the observed \ion{O}{6} and \ion{H}{1} 
column densities in the broad component and the upper 
limits in Table~\ref{lineprop}. Note that since 
$J_{\nu}$(LL) is fixed at an assumed value, varying $U$ is 
tantamount to varying $n_{\rm H}$. Since the 
photoionization model computes the H ion fraction, once $U$ 
is constrained, the total H column density and thickness 
(or pathlength) of the absorber are also constrained.

\begin{figure}
\plotone{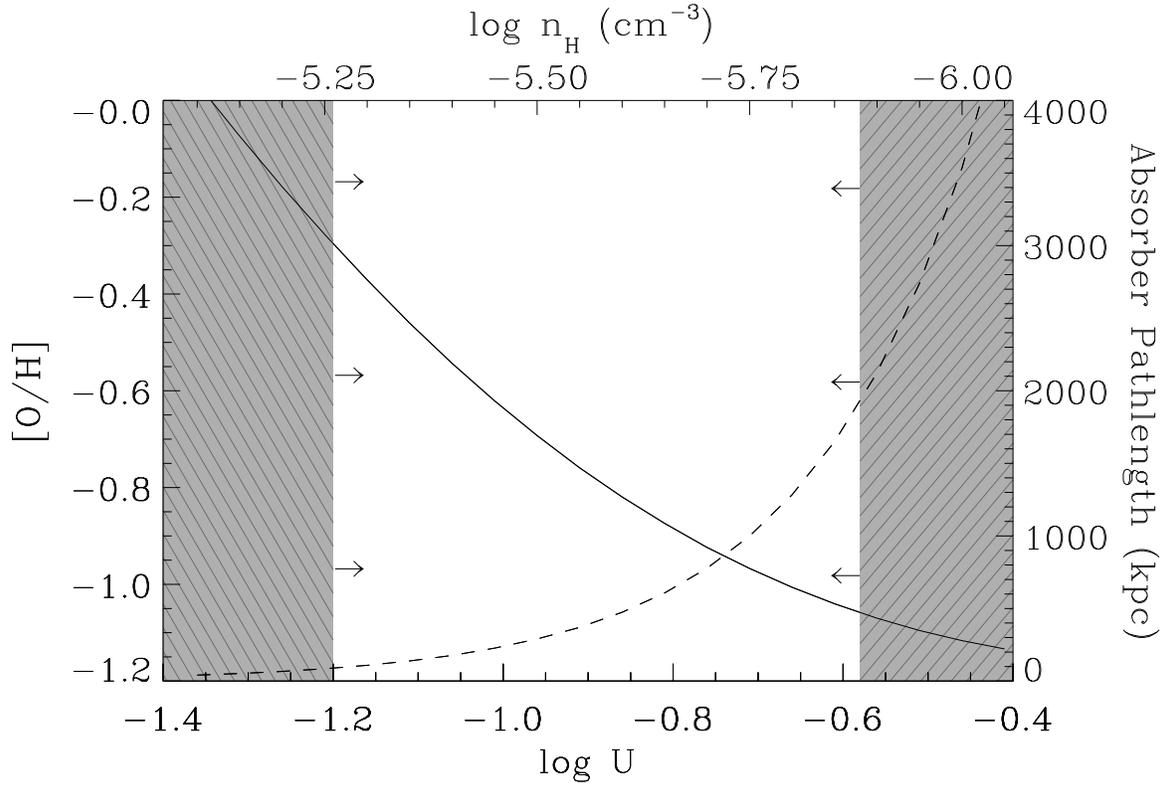}
\caption[]{The logarithmic oxygen abundance [O/H] (solid 
line) required to produce the observed \ion{O}{6} and 
\ion{H}{1} column densities {\it in the broad component} of 
the absorber at \zabs\ = 0.1212, assuming the gas is 
photoionized by the UV background from QSOs, as a function 
of the ionization parameter (bottom axis) and the total H 
density (top axis) for $J_{\nu }$(\ion{H}{1}) = $1 \times 
10^{-23}$ ergs s$^{-1}$ cm$^{-2}$ Hz$^{-1}$ sr$^{-1}$. The 
the \ion{O}{6} and \ion{H}{1} ionization fractions vs. log 
$U$ were calculated with CLOUDY (see text). The dashed line 
shows the required pathlength through the absorber vs. log 
$U$ using the scale on the right axis.\label{photoh}}
\end{figure}

The results of the photoionization models are shown in 
Figure~\ref{photoh} where we have used the $f$(\ion{O}{6}) 
and $f$(\ion{H}{1}) from the photoionization code to derive 
[O/H] vs. log $U$ using equation~\ref{metlim}. From the 
measurements available in Table~\ref{lineprop}, we find 
that the lower limit on $N$(\ion{O}{6})/$N$(\ion{C}{4}) is 
the most useful for constraining the ionization parameter; 
from this ratio we derive log $U \geq -1.20$. A rough upper 
limit on $U$ is provided by the required pathlength through 
the absorber. If the pathlength is too large, the lines 
would be broader than the observed lines due to 
cosmological expansion. Using the rough upper limit on $l$ 
derived in \S 4, we find log $U \lesssim -0.6$ . As before, 
the gray hatching in Figure~\ref{photoh} indicates regions 
which are not consistent with these constraints. From this 
figure, we see that the observational constraints are 
satisfied if $-1.1 \leq$ [O/H] $\lesssim -0.3$ and $85 \leq 
l \lesssim$ 1900 kpc. To derive these metallicities, we 
have assumed log $N$(\ion{H}{1}) = 13.78 from the broad 
component fit shown in Figure 1. In the photoionized 
scenario, the \ion{H}{1} \lya absorption line arising in 
the \ion{O}{6} gas could be substantially narrower, and 
this would lead to a lower \ion{H}{1} column density. This, 
in turn, would require an even higher oxygen abundance to 
produce the observed $N$(\ion{O}{6}) [see eqn. 1]. However, 
log $N$(\ion{H}{1}) = 13.78 is likely a reasonable upper 
limit in this case, and therefore these models provide 
lower limits on the metallicity if the gas is photoionized.

The physical properties implied by the photoionization 
models are reasonable. For example, Schaye (2001) has 
analytically derived expressions for the density and size 
of \lya absorbers, assuming only that the gas is 
photoionized and in local hydrostatic equilibrium. Using 
his equations (8) and (12), and assuming the gas 
temperature $T \sim 10^{4}$K, the \ion{H}{1} 
photoionization rate $\Gamma \sim 10^{-13.3}$ s$^{-1}$, and 
the fraction of the mass in gas $f_{\rm g} \sim 0.2$, we 
obtain log $n_{\rm H} \lesssim$ -5.6 and $l \gtrsim$ 300 
kpc for log $N$(\ion{H}{1}) $\leq$ 13.78. These properties 
are entirely consistent with the properties inferred from 
the CLOUDY models (see Figure~\ref{photoh}). Perhaps the 
most serious concern about these photoionization models is 
that the IGM must be substantially and uniformly enriched 
with metals over a long pathlength. This problem is less 
severe at higher metallicities because the required 
pathlengths are smaller. However, the minimum absorber 
thickness (set by the lower limit on $U$ from 
\ion{O}{6}/\ion{C}{4}) is 85 kpc for [O/H] = $-0.3$. Some 
QSO absorbers have been observed which apparently have very 
low metallicities (e.g., Shull et al. 1999a), so the IGM 
has not been {\it generally} enriched to this level, but 
the system of this paper could arise in a more metal-rich 
region. As noted above, the mean metallicities predicted by 
simulations (e.g., Cen \& Ostriker 1999b; Aguirre et al. 
2001) are only marginally consistent with such a high 
metallicity, but again this could be a particularly high 
$Z$ region. Cen \& Bryan (2001) have argued that most of 
the metals in the diffuse, low-density intergalactic gas 
are provided by small galaxies ($M < 10^{9} M_{\odot}$) and 
are injected into the IGM at $z >$ 4. Because larger 
galaxies which form later do not widely disperse their 
metals, Cen \& Bryan expect that low-z, low column density 
\lya clouds will have low metallicities. A similar picture 
has been favored by Heckman et al. (2000) based on the 
observed properties of supernova-driven ``superwinds'' from 
low$-z$ starburst galaxies and high$-z$ Lyman-break 
galaxies (Pettini et al. 2001). Heckman et al. argue that 
metals injected into the IGM via superwinds, primarily from 
the lower-mass galaxies, could increase the IGM metallicity 
to $\sim$1/6 solar metallicity in the intracluster medium 
as well as the general IGM. However, in most supernova-wind 
enrichment scenarios, the \ion{O}{6} gas should be 
predominantly collisionally ionized. Radiation-pressure 
driven outflows of dust from galaxies (e.g., Aguirre et al. 
2001) could transport heavy elements more gently.

It is interesting to note that Savage et al. (2001) have 
identified an intervening \ion{O}{6} absorber at \zabs\ = 
0.06807 in the spectrum of PG0953+415 which is apparently 
well-explained by photoionization. In this case, 
\ion{H}{1}, \ion{C}{3}, \ion{C}{4}, and \ion{N}{5} are also 
detected and are well-aligned with the \ion{O}{6}, and all 
of the column densities are reasonably matched by a 
photoionization model, albeit with a somewhat uncomfortably 
high metallicity/pathlength combination. Perhaps more 
importantly, the \ion{H}{1} \lya and Ly$\beta$ lines 
associated with the \ion{O}{6} are narrow and appear to be 
composed of only one component, and the upper limit on $T$ 
from the width of the \ion{H}{1} lines is $T \leq 4.1 
\times 10^{4}$ K. This precludes collisional ionization, at 
least in equilibrium, unless the \ion{H}{1} arises in a 
different phase from the \ion{O}{6}. A similar situation is 
found for the \ion{O}{6} system studied by Tripp \& Savage 
(2000), although in this case it is possible to hide a 
broad component in the complicated \lya profile, assuming 
the absorber is a multiphase medium (see their Figure 6). 
While these examples would seem to strongly favor 
photoionization, it is also possible that they represent 
cases where the gas was shock-heated to $\sim 10^{6}$ K and 
then cooled more rapidly than it could recombine, as 
discussed by various authors (e.g., Edgar \& Chevalier 
1986). Table 4 in Tripp \& Savage (2000) summarizes high 
ion ratios predicted by four non-equilibrium collisional 
ionization models. Comparison of the high ion column 
densities measured in this paper to that table shows that 
most non-equilibrium collisional ionization models satisfy 
the observational constraints.  Of course, in this 
hypothesis, the substantial breadth of the \ion{H}{1} 
profile would be unexpected, and some other broadening 
mechanism would need to be invoked. This could simply be 
the kinematics of the gas, e.g., in a galactic wind.

\subsection{X-ray Absorption Lines}

In principle, one promising means to break the collisional 
ionization/photoionization degeneracy is to search for 
X-ray absorption lines, primarily the resonance transitions 
of \ion{O}{7} and \ion{O}{8} at 574 eV and 654 eV, 
respectively (21.6 and 19.0 \AA ). Several groups have 
discussed various aspects of the use of these lines and 
their detection with current X-ray facilities (Shapiro \& 
Bahcall 1980; Aldcroft et al. (1994); Hellsten, Gnedin, \& 
Miralda-Escud\'{e} 1998; Perna \& Loeb 1998; Fang \& 
Canizares 2000; Fang et al. 2001). A complete summary of 
this literature is beyond the scope of this paper, but we 
note that the first attempt to {\it detect} these 
absorption lines in the diffuse IGM (with the {\it Chandra} 
High Energy Transmission Grating Spectrometer) has yielded 
upper limits of $N$(\ion{O}{8}) $\lesssim 10^{17}$ cm$^{-
2}$ (Fang et al. 2001).

\begin{figure}
\plotfiddle{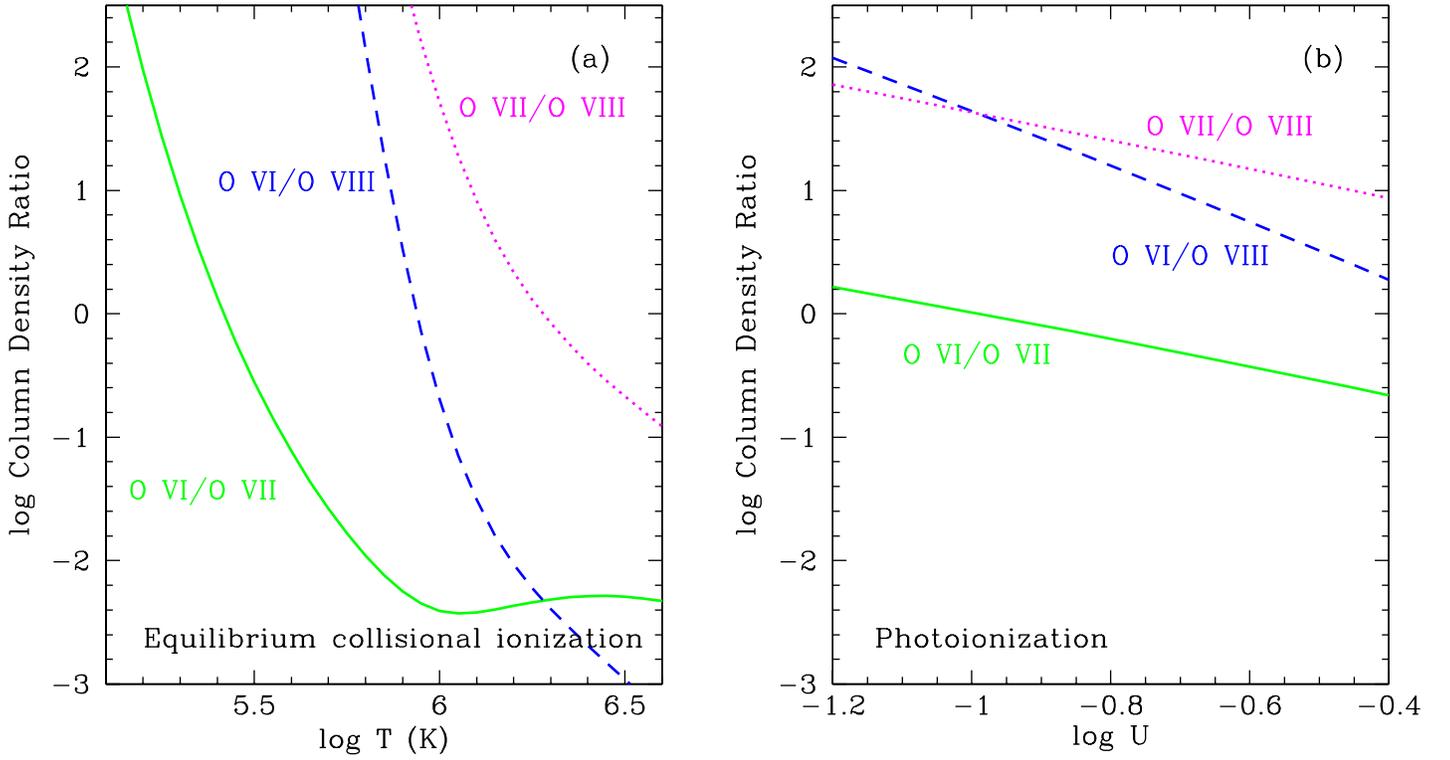}{4in}{270}{75}{75}{-300}{380}
\caption[]{Logarithmic column density ratios predicted for 
\ion{O}{6}/\ion{O}{7} (solid line), \ion{O}{6}/\ion{O}{8} 
(dashed line), and \ion{O}{7}/\ion{O}{8} (dotted line) for 
(a) collisionally ionized gas in equilibrium, and (b) gas 
photoionized by the UV background from QSOs. The ratios in 
the collisionally ionized case are plotted vs. log $T$, 
while the photoionized gas ratios are shown vs. log of the 
ionization parameter.\label{xraylines}}
\end{figure}

Given this rough guideline for detection capabilities with 
current X-ray telescopes, can we hope to determine the 
ionization mechanism of these \ion{O}{6} absorbers by 
searching for the corresponding X-ray absorption lines at 
the same redshifts? To explore this question, we show the 
\ion{O}{6}/\ion{O}{7}, \ion{O}{6}/\ion{O}{8}, and 
\ion{O}{7}/\ion{O}{8} column density ratios predicted for 
collisionally ionized gas (in equlibrium) and photoionized 
gas in Figures~\ref{xraylines}a and \ref{xraylines}b, 
respectively, for the expected ranges of $T$ and $U$. The 
ratios for collisionally ionized gas are based on the 
calculations of Sutherland \& Dopita (1993). The ratios for 
photoionized gas are from the CLOUDY models described in \S 
4.2. Equilibrium collisional ionization and photoionization 
clearly have different signatures in these ratios. For 
example, the \ion{O}{6}/\ion{O}{7} ratio has a much broader 
range in collisionally ionized gas than in photoionized 
gas: given the temperature and ionization parameter 
constraints derived in \S 4 for the particular absorber of 
this paper, the models predict $-1 \lesssim$ log 
(\ion{O}{6}/\ion{O}{7}) $\lesssim 1$ for collisionally 
ionized gas and log (\ion{O}{6}/\ion{O}{7}) $\approx$ 0 for 
photoionized gas. Furthermore, if log 
(\ion{O}{6}/\ion{O}{7}) $\approx$ 0, then the \ion{O}{8} 
line can distinguish between the two models because in this 
situation the \ion{O}{6}/\ion{O}{8} ratio is much larger in 
the collisionally ionized model than the photoionized 
model.\footnote{Along these lines we note that in general, 
the O~VII and O~VIII lines without the O~VI 
constraint may not be able to distinguish between 
collisional ionization and photoionization because the 
dotted lines in Figure~\ref{xraylines} are degenerate, and 
the resolution of current X-ray spectrometers is not 
sufficient to constrain the temperature of the gas based on 
line widths.} While Figure~\ref{xraylines} demonstrates the 
potential of X-ray observations, it unfortunately also 
shows that for the \ion{O}{6} system studied here, the 
\ion{O}{7} and \ion{O}{8} lines are predicted to have 
column densities well below $10^{17}$ cm$^{-2}$ and 
therefore may be difficult or impossible to detect with 
current instruments. X-ray observations of other absorbers 
with larger \ion{O}{6} column densities may be more 
illuminating. However, if the \ion{O}{6} column density is 
too much larger, then the \ion{O}{6} lines will saturate 
making $N$(\ion{O}{6}) difficult to reliably measure. It 
seems that it will be difficult to study \ion{O}{6}, 
\ion{O}{7}, and \ion{O}{8} absorption arising from the same 
gas. If the higher oxygen ions were detected at the 
redshift of one of the H1821+643 \ion{O}{6} systems, this 
would likely indicate the presence of multiphase gas with 
the \ion{O}{7} and \ion{O}{8} revealing a more highly 
ionized phase containing little \ion{O}{6}. 

This discussion has several caveats. First, we have assumed 
that either collisional ionization or photoionization 
dominates. In fact, both ionization mechanisms may 
contribute. Hellsten et al. (1998) argue that 
photoionization by the X-ray background dominates in the 
production of \ion{O}{7} and \ion{O}{8}, even if the gas 
has a temperature of $10^{5} - 10^{7}$ K, but this depends 
on the density of the gas. The \ion{O}{6} might also be 
produced by the combined effects of collisional and 
photoionization. Second, the gas is assumed to be in 
ionization equilibrium. We have already noted that this 
assumption might not hold for collisionally ionized gas. 
Photoionized \ion{O}{6} is likely to be in equilibrium, but 
\ion{O}{7} and \ion{O}{8} produced by photoionization might 
be out of equilibrium because the ionization time scales 
are comparable to the Hubble time (see, e.g., Hellsten et 
al. 1998). Again, it would be valuable to evaulate the 
ratios predicted for these ions in non-equilibrium models 
as well. Finally, the intensity of the X-ray background is 
uncertain. A full assessment of these issues is beyond the 
scope of this paper.

\section{Summary and Discussion}

High-resolution STIS and {\it FUSE} UV spectra of H1821+643 
reveal that the absorption system at \zabs\ = 0.1212 has a 
number of interesting properties including (1) detection of 
\ion{O}{6} and \ion{H}{1} absorption without significant 
detection of any other species, (2) complex \ion{H}{1} 
absorption including an apparently broad component with $b 
\approx$ 85 \kms\ that is aligned with the \ion{O}{6} 
absorption, and (3) association with one or more galaxies 
close to the sight line. Using constraints from these 
spectra, we have examined the ionization mechanism in this 
absorber. We find that equilibrium collisional ionization 
is viable with 5.3 $\leq$ log $T \leq$ 5.6 and $-1.8 \leq$ 
[O/H] $\leq -0.6$. However, the absorption line properties 
can also be explained by photoionization if 
$-1.1 \leq$ [O/H] $\leq -0.3$ and $85 \leq l \lesssim 1900$ 
kpc. In addition, it is possible that the gas is 
collisionally ionized but is not in equilibrium (e.g., 
Edgar \& Chevalier 1986).

There are a number of possible sites in which \ion{O}{6} 
absorption lines could arise in the spectra of low$-z$ 
QSOs. In some cases, \ion{O}{6} absorption likely occurs in 
the immediate environment of the QSO itself (e.g., Savage 
et al. 1998; Papovich et al. 2000; Ganguly et al. 2001, and 
references therein). The large displacement from the QSO 
redshift and the close association with galaxies near the 
line of sight indicates that the absorber studied in this 
paper is {\it not} in this class. Rather, it is an 
intervening absorber. However, the detailed nature of the 
intervening absorption is still unclear -- it could occur 
in the bound ISM of an individual galaxy, the escaping gas 
in a galactic wind, the intragroup medium of a galaxy 
group, or a more remote region of a large scale structure 
or the diffuse IGM.

The galactic wind hypothesis is appealing because 
\ion{O}{6} absorption has been detected in such an outflow 
from the nearby starburst galaxy NGC1705 (Heckman et al. 
2001). Gas in the temperature range needed to produce 
\ion{O}{6} by collisional ionization is predicted in 
hydrodynamic simulations of starburst galaxy winds (e.g., 
Mac Low \& Ferrara 1999; Strickland \& Stevens 2000). 
However, unlike the \ion{O}{6} absorber studied in this 
paper, the outflow from NGC1705 shows strong absorption 
lines due to low and intermediate ionization stages as well 
as \ion{O}{6}. This absorption by low ions is consistent 
with the hydrodynamic models in which the 
\ion{O}{6}-bearing gas arises in interfaces between cool, 
denser gas and much hotter gas (see Figure 4c in Heckman et 
al. 2001). In this model, low ion absorption occurs in the 
cool gas, \ion{O}{6} exists in the interface, and the 
hotter gas produces X-ray emission. The NGC1705 
observations probe the ISM and outflow close to the central 
starburst and its host galaxy, but the \zabs\ = 0.1212 
\ion{O}{6} system is at a substantially larger projected 
distance from any galaxy known to be at the absorber 
redshift. One could speculate that the character of 
absorption from a wind is different at larger distances 
away from the star cluster, but these winds have not been 
observationally or theoretically studied at such large 
impact parameters.

The hypothesis that the absorption occurs in the ISM of a 
galaxy like the Milky Way suffers similar criticisms. While 
\ion{O}{6} is commonly detected in the disk and halo of the 
Milky Way (Jenkins 1978; Savage et al. 2000), it would 
typically be accompanied by absorption from other heavy 
elements. Galactic high velocity clouds, including clouds 
stripped out of satellite galaxies, also frequently show 
associated \ion{O}{6} (Sembach et al. 2000), and high 
velocity clouds have approximately the same metallicity as 
the absorber studied in this paper (Wakker et al. 1999; 
Gibson et al. 2000; Richter et al. 2001). However, a sight 
line through such an entity would have a much higher 
\ion{H}{1} column density and again would be expected to be 
detected in other ions.\footnote{The particular absorber 
studied here is only detected in O~VI and H~I, 
but other intervening O~VI systems have been detected 
in other species (Tripp et al. 2000; Chen \& Prochaska 
2000; Savage et al. 2001).}

Perhaps the most appealing hypothesis is that the \zabs\ = 
0.1212 absorber is due to gas between galaxies in a group 
or unvirialized filamentary structure. Diffuse X-ray 
emission indicating the presence of a hot intragroup medium 
has been detected from many galaxy groups (Mulchaey 2000 
and references therein), but mainly groups that are 
elliptical-rich (Mulchaey et al. 1996a; Zabludoff \& 
Mulchaey 1998). Mulchaey et al. (1996b) have suggested that 
spiral-rich galaxy groups also contain hot intragroup gas, 
but the gas is not quite hot enough to be detected in X-ray 
emission ($T \lesssim 4 \times 10^{6}$ K). They predict 
that the \ion{O}{6} absorption lines should be detectable 
in such an intragroup medium. It is interesting that the 
\zabs\ absorber fits this hypothesis in many respects: (1) 
there are several galaxies near the sight line at this 
redshift (see Figure~\ref{lyagal}), (2) at least one of the 
galaxies is a spiral/actively star-forming galaxy, and (3) 
the breadth of the \ion{O}{6} absorption is considerably 
larger than expected from thermal motions (\S 4), which is 
consistent with additional broadening due to the velocity 
dispersion of the group.\footnote{The typical velocity 
dispersion of a spiral-rich group is $\sim 100$ \kms\ 
(Mulchaey et al. 1996b).} However, this may not be a bound 
group. Given our currently limited information, it is 
difficult to derive additional information about the 
collection of galaxies near this \ion{O}{6} absorber, but 
we do note that the spatial extent of the ensemble appears 
to be larger than that of a typical poor group (c.f., 
Zabludoff \& Mulchaey 1998).  It would be valuable to 
obtain additional galaxy redshift measurements and 
high-resolution images for morphological classification.

In \S 1 we outlined several areas in which additional 
effort is needed to elucidate the nature of the \ion{O}{6} 
systems and their contribution to the baryon budget. The 
main goal of this paper has been to scrutinize the 
ionization mechanism to test the idea that a substantial 
quantity of shock-heated hot gas is present in the IGM at 
the current epoch. We find that the \ion{O}{6} absorber at 
\zabs\ = 0.1212 in the spectrum of H1821+643 has properties 
consistent with collisionally ionized hot gas such as a 
broad \ion{H}{1} component aligned with the \ion{O}{6} and 
a high \ion{O}{6}/\ion{C}{4} ratio. This is encouraging 
evidence in support of the shock-heated IGM hypothesis. 
However, we cannot rule out photoionization or a 
combination of collisional and photoionization processes. 
The degree of IGM metal enrichment required by photoionization 
exceeds the predictions 
of some models, but given our limited understanding of how 
metals are transported out of galaxies and the variety of 
possibilities for the absorption site, this is not 
sufficient grounds to dismiss photoionization. 
Analyses of other low$-z$ \ion{O}{6} systems (e.g., Tripp 
\& Savage 2000) have reached similar conclusions regarding 
collisional vs. photoionization.

Given the general difficulty encountered in attempts to 
definitively identify the ionization mechanism in 
individual systems, it will be important carry out 
statistical analyses of \ion{O}{6} absorber samples 
including comparisons to various models. For example, Cen 
et al. (2001) and Fang \& Bryan (2001) have recently 
predicted the properties of \ion{O}{6} absorbers based on 
cosmological simulations. The absorber statistics predicted 
by these models [e.g., $dN/dz$ and $\Omega_{\rm 
b}$(\ion{O}{6})] appear to be in reasonable agreement with 
current observations. In these simulations, \ion{O}{6} 
absorption arises in both collisionally ionized and 
photoionized gas. However, the photoionized systems are 
narrower and have lower equivalent widths than the 
collisionally ionized absorbers. Tripp (2002) has recently 
presented the $b-$values and column densities of twenty 
low$-z$ \ion{O}{6} lines observed with STIS and {\it FUSE} 
(see his Figure 5). It is interesting to note that while 
there are a few apparently narrow \ion{O}{6} lines in the 
observations compiled by Tripp (2002), the majority of the 
\ion{O}{6} lines in that sample have 
$b-$values consistent with an origin in hot gas. The 
$b-$value of an \ion{O}{6} line in gas at $T \sim 300,000$ 
K is $\sim$18 \kms , and the median $b-$value of the 
observed sample is 22 \kms . In addition, the apparently 
narrow \ion{O}{6} lines tend to be weaker systems as well, 
as predicted by the simulations. These preliminary results 
appear to be fully consistent with the cosmological 
simulations. We look forward to similar analyses using 
larger observational samples and a variety of models. 

\acknowledgements

We thank Anthony Aguirre and Joop Schaye for helpful 
discussions and comments. This research has made use of 
software developed by the STIS Team for the reduction of 
STIS data, and we thank the STIS Team for access to this 
software. The coding of the Robertson optimal extraction 
was carried out by Bob Hill, and we greatly appreciate this 
effort. We acknowledge support from NASA through grants 
GO-08165.01-97A and GO-08182.01-98A from the Space 
Telescope Science Institute as well as NASA Astrophysical 
Theory grant NAG5-7262. This work is also based on data 
obtained for the Guaranteed Time Team by the NASA-CNES-CSA 
{\it FUSE} mission operated by the Johns Hopkins 
University. Financial support to US participants has been 
provided by NASA contract NAS5-32985.


\begin{references}
\footnotesize
\reference{} Aguirre, A., Hernquist, L., Schaye, J., Katz, 
N., Weinberg, D. H., \& Gardner, J. 2001, ApJ, in press 
(astro-ph/0105065)
\reference{} Aldcroft, T., Elvis, M., McDowell, J., \& 
Fiore, F. 1994, ApJ, 437, 584
\reference{} Allende Prieto, C., Lambert, D. L., \& 
Asplund, M. 2001, ApJ, 556, L63
\reference{} Bowen, D. V., Pettini, M., \& Boyle, B. J. 
1998, MNRAS, 297, 239
\reference{} Bowers, C. A., et al. 2001, in preparation
\reference{} Brown, T. M., Kimble, R. A., Ferguson, H. C., 
Gardner, J. P., Collins, N. R., \& Hill, R. S. 2000, AJ, 
120, 1153
\reference{} Burles, S., \& Tytler, D. 1998, ApJ, 499, 699
\reference{} Cen, R., \& Bryan, G. L. 2001, ApJ, 546, L81
\reference{} Cen, R., \& Ostriker, J. P. 1999a, ApJ, 514, 1
\reference{} Cen, R., \& Ostriker, J. P. 1999b, ApJ, 519, 
L109
\reference{} Cen, R., Tripp, T. M., Ostriker, J. P., \& 
Jenkins, E. B. 2001, ApJ, in press (astro-ph/0106204)
\reference{} Chen, H.-W., \& Prochaska, J. X. 2000, ApJ, 
543, L9
\reference{} Croft, R. A. C., Di Matteo, T., Dav\'{e}, R., 
Hernquist, L., Katz, N., Fardal, M. A., \& Weinberg, D. H. 
2001, ApJ, 557, 67
\reference{} Dav\'{e}, R., Hernquist, L., Katz, N., \& 
Weinberg, D. H. 1999, ApJ, 511, 521
\reference{} Dav\'{e}, R., et al. 2001, ApJ, 552, 473
\reference{} Dav\'{e}, R., \& Tripp, T. M. 2001, ApJ, 553, 
528
\reference{} Davis, D. S., Mulchaey, J. S., \& Mushotzky, 
R. F. 1999, ApJ, 511, 34
\reference{} Donahue, M., Aldering, G., \& Stocke, J. T. 
1995, ApJ, 450, L45
\reference{} Draine, B. T., \& McKee, C. F. 1993, ARA\&A, 
31, 373
\reference{} Edgar, R. J., \& Chevalier, R. A. 1986, ApJ, 
310, L27
\reference{} Fang, T., \& Bryan, G. L., ApJ, submitted
\reference{} Fang, T., \& Canizares, C. R. 2000, ApJ, 539, 
532
\reference{} Fang, T., Marshall, H. L., Bryan, G. L., 
Canizares, C. R. 2001, ApJ, 555, 356
\reference{} Fardal, M., Giroux, M. L., \& Shull, J. M. 
1998, AJ, 115, 2206
\reference{} Ferland, G. J., Korista, K. T., Verner, D. A., Ferguson, 
J. W., Kingdon, J. B., \& Verner, E. M. 1998, PASP, 110, 761
\reference{} Fitzpatrick, E. L., \& Spitzer, L. 1997, ApJ, 475, 623
\reference{} Fukugita, M., Hogan, C. J., \& Peebles, P. J. E. 1998, 
ApJ, 503, 518
\reference{} Ganguly, R., Bond, N. A., Charlton, J. C., Eracleous, M., 
Brandt, W. N., \& Churchill, C. W. 2001, ApJ, 549, 133
\reference{} Gibson, B. K., Giroux, M. L., Penton, S. V., Putman, M. 
E., Stocke, J. T., \& Shull, J. M. 2000, AJ, 120, 1830
\reference{} Grevesse, N., \& Anders, E. 1989, in AIP Conf. Proc. 183, 
Cosmic Abundances of Matter, ed. C. J. Waddington (New York: AIP), 1
\reference{} Grevesse, N., \& Noels, A. 1993, in Origin and Evolution 
of the Elements, ed. N. Prantzos, E. Vangioni-Flam, \& M. Cass\'{e} 
(Cambridge: Cambridge University Press), 15
\reference{} Haardt, F., \& Madau, P. 1996, ApJ, 461, 20
\reference{} Hamann, F., \& Ferland, G. 1999, ARA\&A, 37, 487
\reference{} Heckman, T. M., Lehnert, M. D., Strickland, D. K., \& 
Armus, L. 2000, ApJS, 129, 493
\reference{} Heckman, T. M., Sembach, K. R., Meurer, G. R., Strickland, 
D. K., Martin, C. L., Calzetti, D., \& Leitherer, C. 2001, ApJ, 554, 
1021
\reference{} Hellsten, U., Gnedin, N. Y., \& Miralda-Escud\'{e} 1998, 
ApJ, 509, 56
\reference{} Holweger, H. 2001, in Solar and Galactic Composition, ed. 
R. F. Wimmer-Schweingruber (Berlin: Springer), in press
\reference{} Hwang, U., Mushotzky, R. F., Burns, J. O., Fukazawa, Y., 
\& White, R. A. 1999, ApJ, 516, 604
\reference{} Jenkins, E. B. 1978, ApJ, 219, 845
\reference{} Kerr, F. J., \& Lynden-Bell, D. 1986, MNRAS, 221, 1023
\reference{} Kimble, R. A., et al. 1998, ApJ, 492, L83
\reference{} Kulkarni, V. P., \& Fall, S. M. 1993, ApJ, 413, L63
\reference{} Kuntz, K. D., Snowden, S. L., \& Mushotzky, R. F. 2001, 
ApJ, 548, L119
\reference{} Mac Low, M.-M., \& Ferrara, A. 1999, ApJ, 513, 142
\reference{} Maloney, P. 1993, ApJ, 414, 41
\reference{} McKee, C. F., \& Hollenbach, D. J. 1980, ARA\&A, 18, 219
\reference{} Moos, H. W., et al. 2000, ApJ, 538, L1
\reference{} Morton, D. C. 1991, ApJS, 77, 119
\reference{} Morton, D. C. 2001, in preparation
\reference{} Mulchaey, J. S. 2000, ARA\&A, 38, 289
\reference{} Mulchaey, J. S., Davis, D. S., Mushotzky, R. 
F., \& Burstein, D. 1996a, ApJ, 456, 80
\reference{} Mulchaey, J. S., Mushotzky, R. F., Burstein, 
D., Davis, D. S. 1996b, ApJ, 456, L5
\reference{} Oegerle, W. R., Tripp, T. M., Sembach, K. R., 
Jenkins, E. B., Bowen, D. V., Cowie, L. L., Green, R. F., 
Kruk, J. W., Savage, B. D., Shull, J. M., \& York, D. G. 
2000, ApJ, 538, L23
\reference{} Papovich, C., Norman, C., Bowen, D. V., 
Heckman, T., Savaglio, S., Koekemoer, A. M., \& Blades, J. 
C. 2000, ApJ, 531, 654
\reference{} Penton, S. V., Shull, J. M., \& Stocke, J. T. 
2000, ApJ, 544, 150
\reference{} Perna, P., \& Loeb, A. 1998, ApJ, 503, L135
\reference{} Pettini, M., Shapley, A. E., Steidel, C. C., 
Cuby, J.-G., Dickinson, M., Moorwood, A. F. M., Adelberger, 
K. L., \& Giavalisco, M. 2001, ApJ, 554, 981
\reference{} Phillips, L. A., Ostriker, J. P., \& Cen, R. 
2001, ApJ, 554, L9
\reference{} Persic, M., \& Salucci, P. 1992, MNRAS, 258, 
14P
\reference{} Rauch, M., Miralda-Escud\'e, J., Sargent, W.L.W., Barlow, 
T.A., Hernquist, L., Weinberg D.H., Katz, N., Cen, R., Ostriker, J.P. 
1997b, ApJ, 489, 7
\reference{} Richter, P., Sembach, K. R., Wakker, B. P., Savage, B. D., 
Tripp, T. M., Murphy, E. B., Kalberla, P. M. W., \& Jenkins, E. B. 
2001, ApJ, in press (astro-ph/0105466)
\reference{} Rines, K., Mahdavi, A., Geller, M. J., Diaferio, A., Mohr, 
J. J., \& Wegner, G. 2001, ApJ, 555, 558
\reference{} Robertson, J. G. 1986, PASP, 98, 2000
\reference{} Sahnow, D. J., et al. 2000, ApJ, 538, L7
\reference{} Savage, B. D., et al. 2000, ApJ, 538, L27
\reference{} Savage, B. D., Sembach, K. R., Tripp, T. M., 
\& Richter, P. 2001, ApJ, submitted
\reference{} Savage, B. D., \& Sembach, K. R. 1991, ApJ, 
379, 245
\reference{} Savage, B. D., Tripp, T. M., \& Lu, L. 1998, 
AJ, 115, 436
\reference{} Scharf, C., Donahue, M., Voit, G. M., Rosati, 
P., \& Postman, M. 2000, 528, L73
\reference{} Schaye, J. 2001, ApJ, in press (astro-
ph/0104272)
\reference{} Sembach, K. R., et al. 2000, ApJ, 538, L31
\reference{} Sembach, K. R., \& Savage, B. D. 1992, ApJS, 
83, 147
\reference{} Shapiro, P. R., \& Bahcall, J. N. 1980, ApJ, 
241, 1
\reference{} Shull, J. M., Penton, S. V., Stocke, J. T., 
Giroux, M. L., van Gorkom, J. H., Lee, Y. H., \& Carilli, 
C. 1999a, AJ, 116, 2094
\reference{} Shull, J. M., Roberts, D., Giroux, M. L., 
Penton, S. V., \& Fardal, M. A, 1999b, AJ, 118, 1450
\reference{} Shull, J. M., Stocke, J. T., \& Penton, S. V. 
1996, AJ, 111, 72
\reference{} Sofia, U. J., \& Meyer, D. M. 2001, ApJ, 554, 
L221
\reference{} Strickland, D. K., \& Stevens, I. R. 2000, 
MNRAS, 314, 511
\reference{} Sutherland, R. S., \& Dopita, M. A. 1993, 
ApJS, 88, 253
\reference{} Tripp, T. M. 2002, in ASP Conf. Ser. 
Extragalactic Gas at Low Redshift, A Workshop in Honor of 
Ray Weymann, ed. J. S. Mulchaey \& J. Stocke (San 
Francisco: ASP), in press (astro-ph/0108278)
\reference{} Tripp, T. M., Lu, L., \& Savage, B. D. 1998, 
ApJ, 508, 200
\reference{} Tripp, T. M., \& Savage, B. D. 2000, ApJ, 542, 
42
\reference{} Tripp, T. M., Savage, B. D., \& Jenkins, E. B. 
2000, 534, L1
\reference{tum99} Tumlinson, J., Giroux, M. L., Shull, J. M., \& 
Stocke, J. T. 1999, AJ, 118, 2148
\reference{} Verner, D. A., Tytler, D., \& Barthel, P. D. 
1994, ApJ, 430, 186
\reference{} Vogel, S., Weymann, R., Rauch, M., \& 
Hamilton, T. 1995, ApJ, 441, 162
\reference{} Voit, G. M., Evrard, A. E., \& Bryan, G. L. 
2001, ApJ, 548, L123
\reference{} Wakker,B. P., et al. 1999, Nature, 402, 388
\reference{} Weinberg, D.H., Hernquist, L., Miralda-Escud\'{e}, J., \& 
Katz, N., 1997, ApJ, 490, 564
\reference{} Woodgate, B. E., et al. 1998, PASP, 110, 1183
\reference{} Zabludoff, A. I., \& Mulchaey, J. S. 1998, 
ApJ, 496, 39
\end{references}
\end{document}